\RequirePackage{rotating}
\documentclass[a4paper,fleqn,usenatbib]{mnras}

\usepackage[T1]{fontenc}
\usepackage{ae,aecompl}

\usepackage{graphicx}	% Including figure files
\usepackage{amsmath}	% Advanced maths commands
\usepackage{amssymb}	% Extra maths symbols
\usepackage{subfigure}
\usepackage{times}
\usepackage{multirow,booktabs,colortbl,tabularx}
\usepackage{placeins}
\usepackage{tikz}
\usepackage[bookmarks=false]{hyperref}
\usepackage{booktabs}
\usepackage{array}
\newcolumntype{P}[1]{>{\centering\arraybackslash}p{#1}}
\usepackage{dblfloatfix}
\usepackage{rotating}
\usepackage{cleveref}
\usepackage{caption}

\title[Very Hard LMXBs]{Very hard states in neutron star low-mass X-ray binaries}

\author[A.S. Parikh et al.]
{\parbox{\textwidth}{
A.S. Parikh$^{1}$\thanks{E-mail: a.s.parikh@uva.nl},
R. Wijnands$^{1}$,
N. Degenaar$^{1}$,
D. Altamirano$^{2}$, 
A. Patruno$^{3}$, 
N.V. Gusinskaia$^{1}$,
J.W.T. Hessels$^{1,4}$
}
\vspace{0.4cm}\
\\
$^{1}$Anton Pannekoek Institute for Astronomy, University of Amsterdam, Postbus 94249, 1090 GE Amsterdam, The Netherlands\\
$^{2}$Department of Physics and Astronomy, Southampton University, Southampton SO17 1BJ, UK\\
$^{3}$Leiden Observatory, Leiden University, Postbus 9513, 2300 RA, Leiden, The Netherlands\\
$^{4}$ASTRON, the Netherlands Institute for Radio Astronomy, Postbus 2, 7990 AA, Dwingeloo, The Netherlands
}

\date{Accepted XXX. Received YYY; in original form ZZZ}

\pubyear{2016}

\begin{document}
\label{firstpage}
\pagerange{\pageref{firstpage}--\pageref{lastpage}}
\maketitle

\begin{abstract}
We report on unusually very hard spectral states in three confirmed neutron-star low-mass X-ray binaries (1RXS J180408.9$-$342058, EXO 1745$-$248, and IGR J18245$-$2452) at a luminosity between $\sim 10^{36-37}$ erg s$^{-1}$. When fitting the {\it Swift} X-ray spectra (0.5 -- 10 keV) in those states with an absorbed power-law model, we found photon indices of $\Gamma \sim 1$, significantly lower than the $\Gamma$ = 1.5 -- 2.0 typically seen when such systems are in their so called hard state. For individual sources very hard spectra were already previously identified but here we show for the first time that likely our sources were in a distinct spectral state (i.e., different from the hard state) when they exhibited such very hard spectra. It is unclear how such very hard spectra can be formed; if the emission mechanism is similar to that operating in their hard states (i.e., up-scattering of soft photons due to hot electrons) then the electrons should have higher temperatures or a higher optical depth in the very hard state compared to those observed in the hard state. By using our obtained $\Gamma$ as a tracer for the spectral evolution with luminosity, we have compared our results with those obtained by \citet{wijnands2015low}. We confirm their general results in that also our sample of sources follow the same track as the other neutron star systems, although we do not find that the accreting millisecond pulsars are systematically harder than the non-pulsating systems.

\end{abstract}


%%%%%%%%%%%%%%%%%%%%%%%%%%%%%%%%%%%%%%%%%%%%%%%%%%

%%%%%%%%%%%%%%%%% BODY OF PAPER %%%%%%%%%%%%%%%%%%

\section{Introduction}
Low-mass X-ray binaries (LMXBs) have a compact object, a neutron star (NS) or a black hole (BH), as the primary object, and a low-mass donor star ($\lesssim$1 $M_{\odot}$). The donor star facilitates accretion onto the compact object by overflowing its Roche lobe. Transient LMXBs undergo outbursts lasting weeks to years with outburst X-ray luminosities of $L_\text{X}$ $\sim$ 10$^{35 - 38}$ erg s$^{-1}$, amidst periods of quiescence (with $L_\text{X}$ $\lesssim$ 10$^{34}$ erg s$^{-1}$) that last months to decades.

Many LMXBs are observed to have spectra that become softer with decreasing luminosity below $L_\text{X} \lesssim 10^{36}$ erg s$^{-1}$ \citep[e.g., ][]{padilla2011x,reynolds2014quiescent}.   This can be studied by fitting a phenomenological power-law model to the spectra and using the photon index $\Gamma$ to trace the spectral evolution. The associated $\Gamma$s display an anti-correlation with $L_\text{X}$ in the 0.5 -- 10 keV band. \citet{wijnands2015low} assembled a sample of sources for which they plotted $\Gamma$ against $L_\text{X}$. NSs soften with decreasing luminosities below $L_\text{X}$ $\lesssim$ 10$^{36}$ erg s$^{-1}$ (with a typical $\Gamma$ of $\sim 1.8$) down to $L_\text{X}$ $\sim$ 10$^{34}$ erg s$^{-1}$ ($\Gamma$ $\sim$ 3).  In contrast, BHs are observed to soften only from $\Gamma \sim 1.5$ at around $L_\text{X} \sim 10^{34}$ erg s$^{-1}$ increasing to about $\Gamma \sim 2$ at $L_\text{X} \sim 10^{33}$ erg s$^{-1}$ without further softening at lower $L_\text{X}$. Because of this different behaviour BHs and NSs describe two separate tracks in the $\Gamma$ versus $L_\text{X}$ diagram, although this needs confirmation by studying more sources.

It is typically observed (and therefore commonly assumed) that when the 0.5 -- 10 keV spectra of NS LMXBs are fit with a power-law model the photon index can only be as low as $\Gamma \sim$ 1.5 -- 2.0 (e.g., \citeauthor{lewis2010double} \citeyear{lewis2010double}; \citeauthor{degenaar2012four} \citeyear{degenaar2012four}; \citeauthor{bahramian2013discovery} \citeyear{bahramian2013discovery}; \citeauthor{wijnands2015low} \citeyear{wijnands2015low}). However, in our recent paper \citep{parikh2016potential}, we studied the transient LMXB 1RXS J180408.9$-$342058. During this analysis we found that at the beginning of its 2015 outburst the source displayed very hard spectra with photon indices of $\Gamma$ $\sim$ 1 (in the energy range 0.5 -- 10 keV; in the rest of the paper we will always assume this energy range for the determination of $\Gamma$). This is much harder than expected. Here we study several sources with similar very hard spectra to confirm the existence of such a very hard state in multiple NS systems.

\section{Source Selection and Data Analysis}
\label{sect_data_analysis}
The very hard spectra of 1RXS J180408.9$-$342058 prompted us to search the literature for more sources that may also display such spectral hardness. The recent paper by \citet{tetarenko2016disc} reported that EXO 1745$-$248 showed unusual very hard spectra during the beginning of its 2015 outburst. In addition, we also found that IGR J18245$-$2452 and SAX J1748.9$-$2021 were reported to be similarly very hard \citep{linares2014neutron,ferrigno2014hiccup,bozzo2015swift}. Finally, \citet{del2014puzzling} proposed a tidal disruption event by a planet onto a white dwarf for IGR J17361$-$4441 but \citet{wijnands2015low} put forth an LMXB nature. Since a NS accretor is not firmly ruled out, we included it in our sample.

Using these, we noted that two of the systems (IGR J18245$-$2452 and SAX J1748.9$-$2021) are accreting millisecond X-ray pulsars \citep[AMXPs;][]{altamirano2008intermittent,papitto2013swings}. It was noted by \citet[][although based on limited amount of data]{wijnands2015low} that AMXPs might, at similar $L_\text{X}$, show sightly harder spectra than non-pulsating NS systems. We considered the possibility that the pulsating nature of those sources (thus the presence of a dynamically important magnetic field) might be related to the very hard spectra of the two AMXPs in our sample. To test this hypothesis, we also included the canonical AMXP SAX J1808.4$-$3658 in our study. This AMXP was chosen as it has been extensively monitored by {\it Swift}/XRT (the instrument we use in our study; see below; \citeauthor{patruno2016radio} \citeyear{patruno2016radio}).  The other five sources have also been well monitored using this instrument. In case multiple outbursts were observed for a given source, only the well-sampled outbursts were considered (i.e. the evolution of the outburst was well monitored; based on this criterion we excluded the 2011 outburst of EXO 1745$-$248, the 2011 outburst of SAX J1808.4$-$3658, and the 2012 outburst of 1RXS J180408.9$-$342058). We only include well-sampled outbursts as we wish to track the spectral evolution of the source to ensure that our fit results are not just a statistical fluctuation which could be the case if we consider single sparse pointings.

We followed the spectral evolution of our six sources by fitting a simple absorbed power-law model to all spectra. This allowed us to determine if those sources indeed exhibited very hard spectra but it also allowed us to carry out a follow-up study of \citet{wijnands2015low} to determine if their conclusions still hold when more sources are studied. The \citet{wijnands2015low} data we compare to in this paper corresponds to their Figure 1; we use all the BH data and the NS data that only corresponds to non-pulsating systems with low $N_\text{H}$.

The data were downloaded from the \texttt{HEASARC} archive and were analysed using \texttt{HEASOFT} (version 6.17).  To process the raw data we used \texttt{xrtpipeline}. Circular extraction regions were used to extract the source spectra in \texttt{XSelect}. Depending on the brightness of a given source, we used extraction regions with a radius varying between 25 arcsec and 100 arcsec. We used annular regions to account for the background in both Window Timing (WT) and Photon Counting (PC) mode (varying between 125 arcsec  and 300 arcsec  for the inner radius, and 200 arcsec  and  475 arcsec  for the outer radius). For PC mode observations of sources located in globular clusters (see Table \ref{tab_nh_vals}) the source flux is only a factor of a few above the background caused by other low luminosity sources in the cluster and hence the normal background subtraction method cannot be used. For these observations we extracted spectra from observations when the source was quiescent, using a similar region as when it was active, to serve as the background correction. The backscale keyword was set to correctly scale for different source and background regions\footnote{http://www.swift.ac.uk/analysis/xrt/backscal.php}. The ancillary response files were created using \texttt{xrtmkarf}. The relevant response matrix files for each observation were used. Each spectrum was grouped using \texttt{grppha} -- at least 10 photons per bin for the WT mode and at least 5 photons per bin for the PC mode. All type-I thermonuclear bursts were removed and all the relevant observations (i.e., when the source was relatively bright) were corrected for pile-up.\footnote{http://www.swift.ac.uk/analysis/xrt/pileup.php}

The spectra from the various observations were fitted in \texttt{XS\textsc{pec}} (version 12.9) with an absorbed \texttt{powerlaw} model using W-statistics (background subtracted Cash statistics; \citeauthor{wachter1979parameter} \citeyear{wachter1979parameter}) because of the low number of photons per bin. The \texttt{tbabs} component was used to model the hydrogen column density $N_\text{H}$ using \texttt{VERN} cross-sections and \texttt{WILM} abundances \citep{verner1996atomic,wilms2000absorption}. The value of $N_\text{H}$ used is discussed further in Section \ref{sect_nH}. Due to the low energy spectral residuals in the WT mode, the WT mode data were fit over the 0.7 -- 10 keV range\footnote{http://www.swift.ac.uk/analysis/xrt/digest$\_$cal.php}. The PC data were fit over the 0.5 -- 10 keV range. All the fluxes reported are the unabsorbed fluxes, determined using the convolution model \texttt{cflux} in the 0.5 -- 10 keV range. Luminosities are also given for the 0.5 -- 10 keV range and errors correspond to the 90 per cent confidence range.

\section{Results}
\label{sect_results}

The obtained values of the various parameters resulting from our spectral fits are systematically affected by our assumptions and data reduction. Some of these effects, such as those introduced by distance, pile-up, and fitting a simple model to a more complex spectral shape, have been discussed by \citet{wijnands2015low}. The distances used for the various sources are shown in Table \ref{tab_nh_vals}. \citet{wijnands2015low} also briefly discussed the effect of fitting a simple model to high quality data and the effect of the $N_\text{H}$ value assumed. Here we discuss the effect of the assumed $N_\text{H}$ value in more detail, in relation to our analysis method. %The various sources plotted in \citet{wijnands2015low} may use other cross-sections and abundances. 
Abundances and cross-sections may have a systematic effect on $\Gamma$. Using different abundances changes the absolute value slightly \citep[see Appendix A of][]{plotkin20162015}, however our conclusions do not change when using different cross sections and abundances.

\label{sect_nH}

From our analysis we found that if the $N_\text{H}$ was left free in the spectral fits, the $N_\text{H}$ traced a large range of values (for a given source), especially if the source exhibited different spectral states (e.g., Figure A1 in the Appendix). This change in the $N_\text{H}$ directly affects the $\Gamma$ of the fit as the two components are correlated -- the $\Gamma$ value increases if the $N_\text{H}$ increases \citep[see, e.g.,][]{plotkin2013x}. Since we want to use $\Gamma$ as a tracer for spectral hardness this degeneracy is undesirable. To prevent this issue, we fit the spectra by fixing the $N_\text{H}$ to a particular value, although it was not directly clear to what value we should fix it. In other spectral studies of our sources, the $N_\text{H}$ values used were either the Galactic (line-of-sight extinction) values or the values obtained when $N_\text{H}$ was left free in spectral fits to high quality data. In the end, we analysed our data considering all three approaches: (1) leaving the $N_\text{H}$ free, (2) fixing it to the Galactic $N_\text{H}$, and (3) fixing it to the best-fit $N_\text{H}$. 

For the four sources located in globular clusters (see Table \ref{tab_nh_vals}) the Galactic $N_\text{H}$ was determined using the known reddening E(B $-$ V) or $N_\text{H}$ values previously determined from high quality data were used. If the $N_{\text{H}}$ was calculated from the E(B $-$ V) values the relationship given by \citeauthor{guver2009relation} (\citeyear{guver2009relation}; see also \citeauthor{predehl1995x} \citeyear{predehl1995x}) was used for the conversion. For the remaining two sources (1RXS J180408.9$-$342058 and SAX J1808.4$-$3658), the Galactic $N_\text{H}$ was determined using \citet{dickey1990hi}. The Galactic values of $N_\text{H}$ are shown in Table \ref{tab_nh_vals}.

\begin{table}
\caption{The Galactic and best-fit $N_\text{H}$, and distance used for each source.\textsuperscript{*}}
\label{tab_nh_vals}
\begin{tabular}{p{3cm}p{2.9cm}p{1cm}}
\hline
Source & \hspace{3mm} $N_\text{H}$ (10$^{22}$ cm$^{-2}$)& Distance\tabularnewline
\end{tabular}
\begin{tabular}{p{3cm}>{\centering\arraybackslash}p{1cm}>{\centering\arraybackslash}p{1.5cm}>{\centering\arraybackslash}p{1cm}}
		    &Galactic&Best-fit&(kpc)\tabularnewline
		    \hline
1RXS J180408.9$-$342058 &0.20 &0.41&5.8\tabularnewline
EXO 1745$-$248 & 1.10 & 2.42&5.5 \tabularnewline
IGR J18245$-$2452 & 0.26 & 0.51&5.5 \tabularnewline
SAX J1748.9$-$2021 & 0.57 & 1.41 &8.5\tabularnewline
IGR J17361$-$4441 & 0.25 & 0.26 &13.2\tabularnewline
SAX J1808.4$-$3658 & 0.14 & -\textsuperscript{$\dagger$} &3.5\tabularnewline
\hline
\multicolumn{4}{p{7.8cm}}{\textsuperscript{*}\scriptsize{The sources located in globular clusters are EXO 1745$-$248 (Terzan 5), IGR J18245$-$2452 (M28), SAX J1748.9$-$2021 (NGC 6440), and IGR J17361$-$4441 (NGC 6388). The respective references for the Galactic $N_\text{H}$ values of the cluster sources are \citet{degenaar2012strong}, \citet{harris1996catalog}; E(B $-$ V) = 0.4, \citet{pintore2016broad}, and \citet{bellini2013intriguing}; E(B $-$ V) = 0.37. The source distances are obtained from the following references (given in order) : \citet{chenevez2012integral}, \citet{ortolani2007distances}, \citet[][ using the updated version of 2010]{harris1996catalog}, \citet{ortolani1994low}, \citet{dalessandro2008blue} and \citet{galloway2006helium}.}} \tabularnewline
\multicolumn{4}{p{7.8cm}}{\textsuperscript{$\dagger$}\scriptsize{No best-fit $N_\text{H}$ was calculated for SAX J1808.4$-$3658 (see Appendix A).}} \tabularnewline
\end{tabular}
\end{table}

We determine the best-fit $N_\text{H}$ ourselves. This is done by initially leaving the $N_\text{H}$ free and fitting the various observations with an absorbed power-law model getting various $N_\text{H}$ values. Then we fit a constant to the obtained $N_\text{H}$ values to get the best-fit $N_\text{H}$. If a source is known to show a hard to soft state transition only the data from the hard state is used to determine the constant since it is known that in the hard state a power-law model can typically describe the spectra reasonably adequately (i.e., when the data quality is low) whereas in the soft state the spectra are more complex. Therefore, the $N_\text{H}$ obtained in this way will best resemble the true $N_\text{H}$. The evolution of the free $N_\text{H}$ with time as well as the constant fit (to the relevant observations) is shown in Appendix A where we describe how the best-fit $N_\text{H}$ was determined in more detail. All the calculated best-fit $N_{\text{H}}$ values are shown in Table \ref{tab_nh_vals}. SAX J1808.4$-$3658 (see Appendix A) showed a different $N_\text{H}$ evolution for each of the three outbursts we consider. It also showed $N_\text{H}$ evolution within each outburst. Thus, calculating the best-fit $N_\text{H}$ for this source was not straightforward. Therefore, we have not determined a best-fit $N_\text{H}$ for SAX J1808.4$-$3658. Our values for the other sources are consistent with those reported in the literature \citep{degenaar2012strong,papitto2013swings,linares2014neutron,del2014puzzling,bozzo2015swift,ludlam2016nustar,pintore2016broad,tetarenko2016disc,degenaar2016disk}.

To understand which $N_{\text{H}}$ method to use to compare our targets with the \citet{wijnands2015low} results, we studied the literature references for the data they used to determine how the original papers obtained the spectral fits. We concluded that in (nearly) all results the $N_{\text{H}}$ was obtained by using it as a free parameter in the fit.

\subsection{Photon Index versus Luminosity}

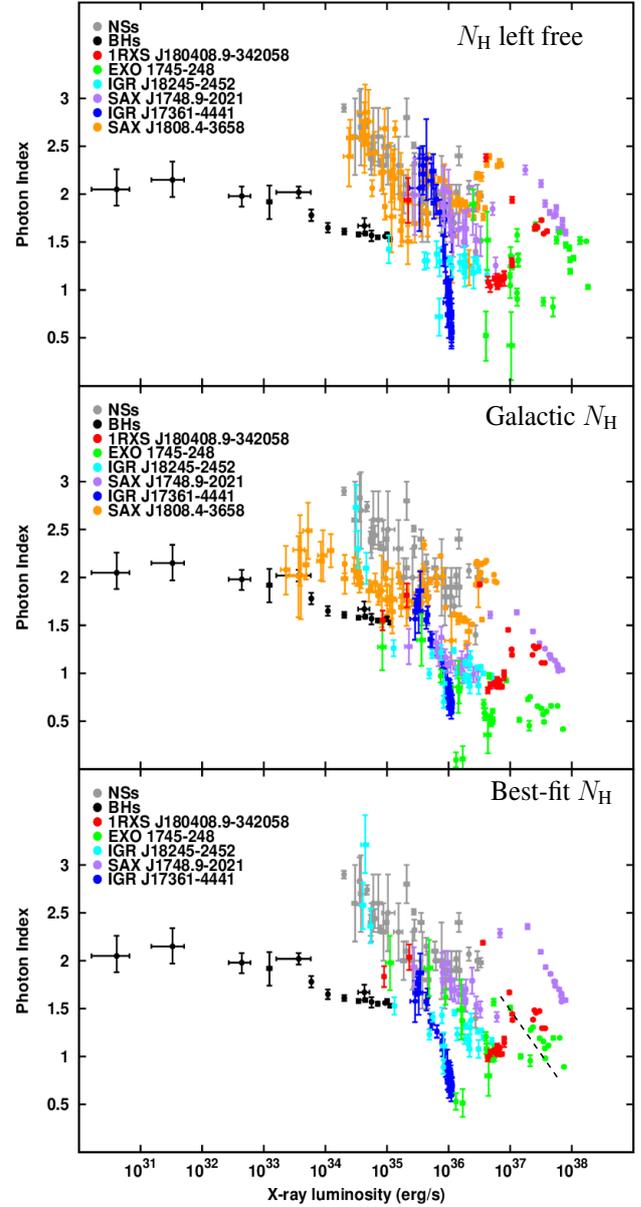
\begin{figure}
\centering
\resizebox{9cm}{17cm}{%
\begin{tikzpicture}
\node[inner sep=0pt] (russell) at (0,0)
    {\includegraphics[scale=0.85]{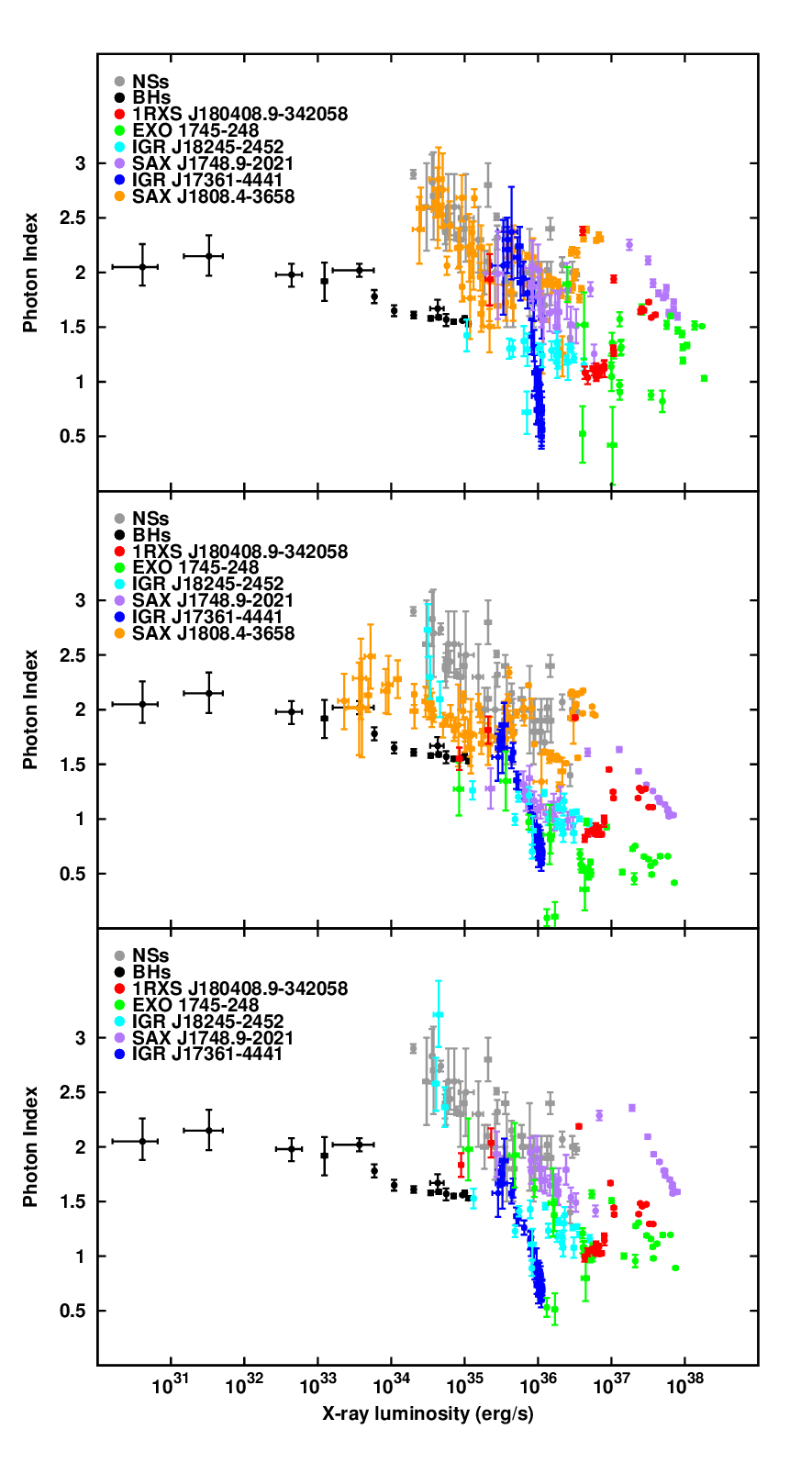}};
\node at (3.4,10) [draw=none]  {\LARGE{$N_\text{H}$ left free}};
\node at (4,3.2) [draw=none]  {\LARGE{Galactic $N_\text{H}$}};
\node at (4,-3.5) [draw=none]  {\LARGE{Best-fit $N_\text{H}$}};
\draw[thick,dashed] (4.1,-8.6) -- (3,-7.1); % 1RXS J1804 : H -> S
%\node at (4.2,-8) [draw=none]  {1};
%\draw[thick,dashed] (2.6,-7.3)--(4.15,-8.6); % EXO 1745 H -> S
%\draw[thick,dashed] (4,-8.5) -- (1.5,-6.5); % 1RXS J1804 : H -> S
%\draw[thick,dashed] (1.5,-6.5) .. controls (2.5,-10) and (2.5,-6) .. (4,-8.6); % EXO 1745 H -> S
%\node at (4.25,-8.6) [draw=none]  {2};
\end{tikzpicture}
}
\caption{The photon index versus luminosity (0.5 -- 10 keV) for various NS (shown in grey) and BH (shown in black) X-ray binaries taken from Figure 1 of \citet{wijnands2015low} along with our targets. The top panel indicates data when $N_\text{H}$ is a free parameter in the fits, the central panel when $N_\text{H}$ is fixed to the Galactic value, and the bottom panel when the $N_\text{H}$ is fixed to the best-fit value. No best-fit $N_\text{H}$ value for SAX J1808.4$-$3658 was calculated (see Appendix A) and therefore the data are not plotted in the lowest panel. The dashed black line in the lowest panel indicates the hard to soft state transition for 1RXS J180408.9$-$342058 and EXO 1745$-$248.}
\label{fig_gamma_v_lumin_main}
\end{figure}

To maintain consistency with \citet{wijnands2015low} we disregarded all fits that have an error >0.5 on $\Gamma$. In the case where we left the $N_{\text{H}}$ free, the errors were systematically larger than in the other two cases, resulting in $\sim$ 10 per cent fewer points (see Figure \ref{fig_gamma_v_lumin_main}). The $\Gamma$ versus $L_\text{X}$ values of our six sources, using different assumptions for the $N_{\text{H}}$ value, are plotted in Figure \ref{fig_gamma_v_lumin_main} along with the results from \citet{wijnands2015low}. The individual $\Gamma$ versus $L_\text{X}$ plots for each source as well as the tabulated values are presented in Figure C1 and Appendix D. We compared our results with previously published results of our sources and we found that they are consistent with those results when similar spectral states were analysed \citep{ferrigno2011swift,linares2014neutron,del2014puzzling,bozzo2015swift,wijnands2015low,tetarenko2016disc}.

The data in the top panel of Figure \ref{fig_gamma_v_lumin_main}, when $N_{\text{H}}$ is left free, indicate the unusually very hard spectra (with $\Gamma \sim 0.5$ -- $1.3$ over the luminosity range $L_\text{X} \sim 10^{36-37}$ erg s$^{-1}$) of our five confirmed NS LMXBs, although SAX J1748.9$-$2021 and SAX J1808.4$-$3658 show only a few data points with very hard spectra and by themselves would not necessarily prove the existence of very hard spectra. However, the combined sample clearly shows the presence of a very hard state. The unclassified source IGR J17361$-$4441 also shows very hard photon indices of $\Gamma \sim 0.5 - 1$ but at slightly lower luminosity $L_\text{X} \sim 10^{36}$ erg s$^{-1}$. Appendix B shows the evolution of $\Gamma$ with time, when $N_{\text{H}}$ is left free. Besides the existence of apparently very hard spectra in our targets, Figure \ref{fig_gamma_v_lumin_main}, top panel, also shows that below $L_\text{X} \lesssim 10^{36}$ erg s$^{-1}$, most of our targets behave in a manner that is in agreement with the results of \citet{wijnands2015low}, although IGR J18245$-$2452 appears to remain very hard between $L_\text{X} \sim 10^{35 - 36}$ erg s$^{-1}$ and for 1RXS J180408.9$-$342058 there is no data available in this range.

The central panel and bottom panel of Figure \ref{fig_gamma_v_lumin_main} shows the data assuming the Galactic $N_{\text{H}}$ and the best-fit $N_{\text{H}}$, respectively. In the middle panel, the $\Gamma$ for all points compared to the free $N_{\text{H}}$ (top panel, Figure \ref{fig_gamma_v_lumin_main}) decreases significantly. Since the Galactic $N_{\text{H}}$ are consistently lower than the free $N_{\text{H}}$ (see Table \ref{tab_nh_vals}) it causes the observed decrease of $\Gamma$. Similarly, a softening of $\Gamma$ is observed when the best-fit $N_{\text{H}}$ is used (lowest panel Figure \ref{fig_gamma_v_lumin_main}) compared to the Galactic $N_{\text{H}}$ as the best-fit $N_{\text{H}}$ tends to be higher.

When assuming the Galactic $N_{\text{H}}$ value in our fits, SAX J1808.4$-$3658 does not fall on the standard NS track but it forms a separate track at significantly lower $\Gamma$s. However, since the data points used by \citet{wijnands2015low} were mostly obtained using a different $N_{\text{H}}$ criteria (see Section \ref{sect_nH}), we cannot directly compare SAX J1808.4$-$3658 and any of of the other sources with this track. Therefore, we will no longer discuss the Galactic $N_{\text{H}}$ results further when comparing our sample to the \citet{wijnands2015low} results.

From Figure \ref{fig_gamma_v_lumin_main}, lower panel, it can be seen that, when using the best-fit $N_\text{H}$, only four of our six sources display very hard spectra with $\Gamma \sim1$. Those sources are 1RXS J180408.9$-$342058, EXO 1745$-$248, IGR J18245$-$2452, and IGR J17361$-$4441. No best-fit $N_\text{H}$ was calculated for SAX J1808.4$-$3658 (see Appendix A) and therefore it has not been plotted. From the different panels in Figure 1, it can be seen that for the AMXP SAX J1748.9$-$2021 the inferred spectral hardness strongly depends on the $N_{\text{H}}$ value used. When using the Galactic $N_{\text{H}}$ in our spectral fits, the source appears to show a very hard spectra ($\Gamma \sim 1$; Fig. \ref{fig_gamma_v_lumin_main}, middle panel). This is consistent with the {\it Swift}/XRT results published by \citet{bozzo2015swift} who assumed this $N_{\text{H}}$. However, when leaving the $N_{\text{H}}$ free (top panel) or fixing it to the best-fit $N_{\text{H}}$, the spectra appear softer and this source is then fully consistent with the standard NS track.

The dashed black line in the lowest panel of Figure \ref{fig_gamma_v_lumin_main} indicates the hard to soft state transition for 1RXS J180408.9$-$342058 and EXO 1745$-$248. The soft state shows the presence of very hard spectra as well, however this is an artifact of using too low $N_\text{H}$ compared to when $N_\text{H}$ is left free  (as seen in Figure A1) and thus we get too low $\Gamma$s (compare bottom with top panel for EXO 1745$-$248). In addition, the soft, thermal component that becomes visible could have temperatures well above a few keV possibly resulting in very hard spectra if one fits with a power-law model in the 0.5 -- 10 keV range (see Section \ref{sect_disc}).

\FloatBarrier
\section{Discussion} 
\label{sect_disc}

We have studied the spectra (using an absorbed power-law model) of six seemingly very hard (candidate) NS LMXBs. The $\Gamma$ values obtained from the spectral fits are strongly dependent on our assumptions about the $N_\text{H}$ values. We find that four sources indeed show very hard spectra down to $\Gamma \sim 1$ (see Fig. \ref{fig_gamma_v_lumin_main}) irrespective of the approach used to determine $N_\text{H}$. The cause of these very hard spectra is unknown. Hard X-rays are expected to be produced by Compton up-scattering of soft photons by hot electrons (e.g., that are present in a corona, an accretion column or a possible boundary layer). The unusual very low $\Gamma$ may be a result of higher electron temperatures or higher optical depths of this Comptonizing medium than those present during the typically observed hard NS state. Our results suggest that we might have identified a new spectral state in NS LMXBs. This conclusion is  strengthened by the different rapid X-ray variability properties we observed during this state compared to the hard state of those sources \citep[see][]{wijnands2017}.

Measurements of the spectra above 10 keV would be useful to investigate the physical process behind such very hard spectra. 1RXS J180408.9$-$342058 is a promising candidate as it has good coverage over a large energy range in several different spectral states. The source has been studied by \citet[][in the energy range 0.45 -- 50 keV]{ludlam2016nustar} when the source was in this very hard state and by \citet[][for 0.7 -- 35 keV]{degenaar2016disk} when it was in the soft state. Comparing the obtained spectra with each other could elucidate the physical mechanism behind the very hard state. However, to do this the spectra have to be reanalysed in a homogenous manner and we are currently performing such reanalysis. The results of this will be reported elsewhere. Here we only wish to stress the existence of a very hard spectral state in NS LMXBs which is important in studies that classify such systems based on spectral hardness alone and this new spectral state needs to be accounted for in models that aim to explain the accretion physics in such systems. In addition, more NS sources need to be studied to determine what fraction of NSs show this very hard state and to determine if some physical property is associated with the presence of such very hard spectra. Also the BH systems need to be studied to check if they can display similar very hard states. If such states are also present in BH systems it would indicate that the physical mechanism in the accretion flow that generates such very hard spectra is not (or only minorly) affected by the presence of a NS surface and/or magnetic field or a BH event horizon. The {\it Swift}/XRT database contains many archival observations that can be used to extend this study for NSs and BHs.

We also compare our data to that reported by \citet{wijnands2015low}. For luminosities below $L_\text{X} \lesssim$ 10$^{36}$ erg s$^{-1}$ all sources but one (IGR J18245$-$2452) line up with the expected NS track when using similar $N_\text{H}$ assumptions as used for the points in \citet[][see Section \ref{sect_nH}]{wijnands2015low}. Although IGR J18245$-$2452 does not follow the NS track at around $\sim 10^{36}$ erg s$^{-1}$, at the lowest luminosities (a few times $10^{34}$ erg s$^{-1}$), the source joins the track (Figure \ref{fig_gamma_v_lumin_main}, bottom; see also \citeauthor{wijnands2015low} \citeyear{wijnands2015low}; those points are absent in Figure \ref{fig_gamma_v_lumin_main} top because the $N_{\text{H}}$ was left free in that plot resulting in such large errors on $\Gamma$ that those point fell outside our selection criteria). If more sources are added to the sample the assumptions on the $N_{\text{H}}$ should be the same as those used by us: i.e., the $N_{\text{H}}$ should ideally be left free but if that is not possible because of the low statistical quality of the data then the $N_{\text{H}}$ should be fixed to the value obtained from higher quality data of the same source.

The absolute value of $\Gamma$ should not be taken at face value since we fit a phenomenological model. The fit parameter $\Gamma$ can only be used to study the broad evolution -- hardening and softening of the spectra of a given source, and to compare different sources with one another. Another limitation is that we fit only the 0.5 -- 10 keV energy range which might mask spectral evolution.
In particular, in the soft state black-body type components are often observed in the spectra and they can have high temperatures (up to several keV). Such spectra below 10 keV still appear as quite hard with relatively low $\Gamma$ (when fitted with a power-law model). This can be clearly seen in Figure \ref{fig_gamma_v_lumin_main} (bottom) where we draw the line between soft and hard state in 1RXS J180408.9$-$342058 and EXO 1745$-$248.

An added complication at the highest luminosities is that during broad band studies using X-ray colors of NS LMXBs (i.e., using hardness-intensity diagrams) it has been found that some sources that accrete at a few tenths of the Eddington rate show the same colors (and hence the same spectral shape) in the soft state at different count rates (and thus different $L_\text{X}$; this is referred to as secular motion; see \citeauthor{homan2010xte} \citeyear{homan2010xte} [i.e., their Figure 1] and \citeauthor{fridriksson2015common} \citeyear{fridriksson2015common} for discussion about this and further references). Although most of these studies were done over a different energy range than the 0.5 -- 10 keV range it is plausible that a similar effect is visible in the 0.5 -- 10 keV energy range. For example, this can be seen in SAX J1808.4$-$3658, although at lower luminosities than typically seen in those other studies. So we urge the reader to be cautious in how to interpret the data points of the sources at their highest $L_\text{X}$. 

IGR J17361$-$4441 is an unusual transient that is not easily classified. Based on its luminosity evolution together with its odd spectra (and how it evolved in time), \citet{del2014puzzling} classified the source as a tidal disruption event of a planet sized body by a white dwarf. Its spectra were odd as they were very hard (with $\Gamma \sim$ 1 and even lower; see also Figure \ref{fig_gamma_v_lumin_main}) and showed the presence of a soft component with very low temperatures ($\sim$ 0.08 keV) that did not change when the luminosity decreased at the end of the outburst. However, we have now found that three confirmed NS transients show similar very hard spectra (1RXS J180408.9$-$342058, EXO 1745$-$248, and IGR J18245$-$2452). The source remains enigmatic (it is the hardest source in our sample, the $\Gamma$ versus $L_\text{X}$ behavior is very steep compared to the other systems, and it is unclear how to explain the constant soft component). However, at low $L_\text{X}$ the source joins the normal NS track for all values of $N_{\text{H}}$ used. Combined with a power spectrum that looks similar to that of accreting NSs or BHs (see appendix of \citeauthor{wijnands2015low} \citeyear{wijnands2015low}; but see \citeauthor{bozzo2014100} \citeyear{bozzo2014100} for an interpretation of these results as a tidal disruption event), this suggest that a NS LMXB nature cannot be discarded for IGR J17361$-$4441.

In our analysis we have studied three AMXPs: SAX J1748.9$-$2021 (an intermittent AMXP), IGR J18245$-$2452 \citep[a transitional millisecond pulsar; a special subset of AMXPs; e.g.,][]{archibald2009radio,papitto2013swings} and SAX J1808.4$-$3658 (the `canonical' AMXP). \citet{linares2014neutron} and \citet{ferrigno2014hiccup} reported on the {\it Swift}/XRT and {\it XMM-Newton} spectra of IGR J18245$-$2452 and noted the hardness of the spectra in this source. They briefly discussed this in the context of the source being a transitional millisecond pulsar and that the NS magnetic field in this system might be related to its very hard spectra. \citet{wijnands2015low} tentatively proposed that AMXPs are systematically harder (in the $L_\text{X}$ range of 10$^{34 - 36}$ erg s$^{-1}$) than non-pulsating NSs, although they highlighted that this hypothesis was only based on limited AMXP data and needed confirmation. In our study we found that indeed IGR  J18245$-$2452 is harder than the non-pulsating systems at $L_\text{X} \sim 10^{36}$ erg s$^{-1}$, although it joins the track at lower $L_\text{X}$. We note that this source was one of the sources used by \citet{wijnands2015low} so it is not surprising that we reproduce their results. However, we found that SAX J1748.9$-$2021 does not show harder than average spectra if the $N_{\text{H}}$ is left free or fixed to the best-fit one (Figure \ref{fig_gamma_v_lumin_main}, top and bottom panels). However, this may be because it only pulsates intermittently and most of the time the source behaves like a non-pulsating NS \citep{altamirano2008intermittent,patruno2009phase}. SAX J1808.4$-$3658 also does not seem to display harder spectra (the caveats for the comparison with \citet{wijnands2015low} should be recalled). In addition, out of the four sources that exhibit very hard spectra -- EXO 1745$-$248, IGR J17361$-$4441, and 1RXS J180408.9$-$342058 are non pulsating sources \citep{wijnands2005hard,bozzo2011igr}. IGR J18245$-$2452 is an AMXP and shows pulsations \citep{papitto2013swings}. This, combined with the fact that not all AMXPs are harder than non pulsating systems, suggests that the hardness of the source spectra does not have a strict connection (if any at all) with the presence of a dynamically important magnetic field.

\section*{Acknowledgements}

AP and RW are supported by a NWO Top Grant, awarded to RW. ND is supported by an NWO Vidi grant. DA acknowledges support from the Royal Society. NVG acknowledges funding from NOVA. JWTH acknowledges funding from an NWO Vidi fellowship and from a European Research Council Starting Grant.



\bibliographystyle{mnras}
 \newcommand{\noop}[1]{}

%%%%%%%%%%%%%%%%%%%%%%%%%%%%%%%%%%%%%%%%%%%%%%%%%%

%%%%%%%%%%%%%%%%% APPENDICES %%%%%%%%%%%%%%%%%%%%%
%%%%%%%%%%%%%%%%% APPENDICES %%%%%%%%%%%%%%%%%%%%%

\appendix
\onecolumn
\section{Best-fit $N_{\text{H}}$ as determined from the individual fits}
\label{sect_best_fit_app}
\setcounter{figure}{0}

\begin{figure}
\centering
\includegraphics[scale=0.9]{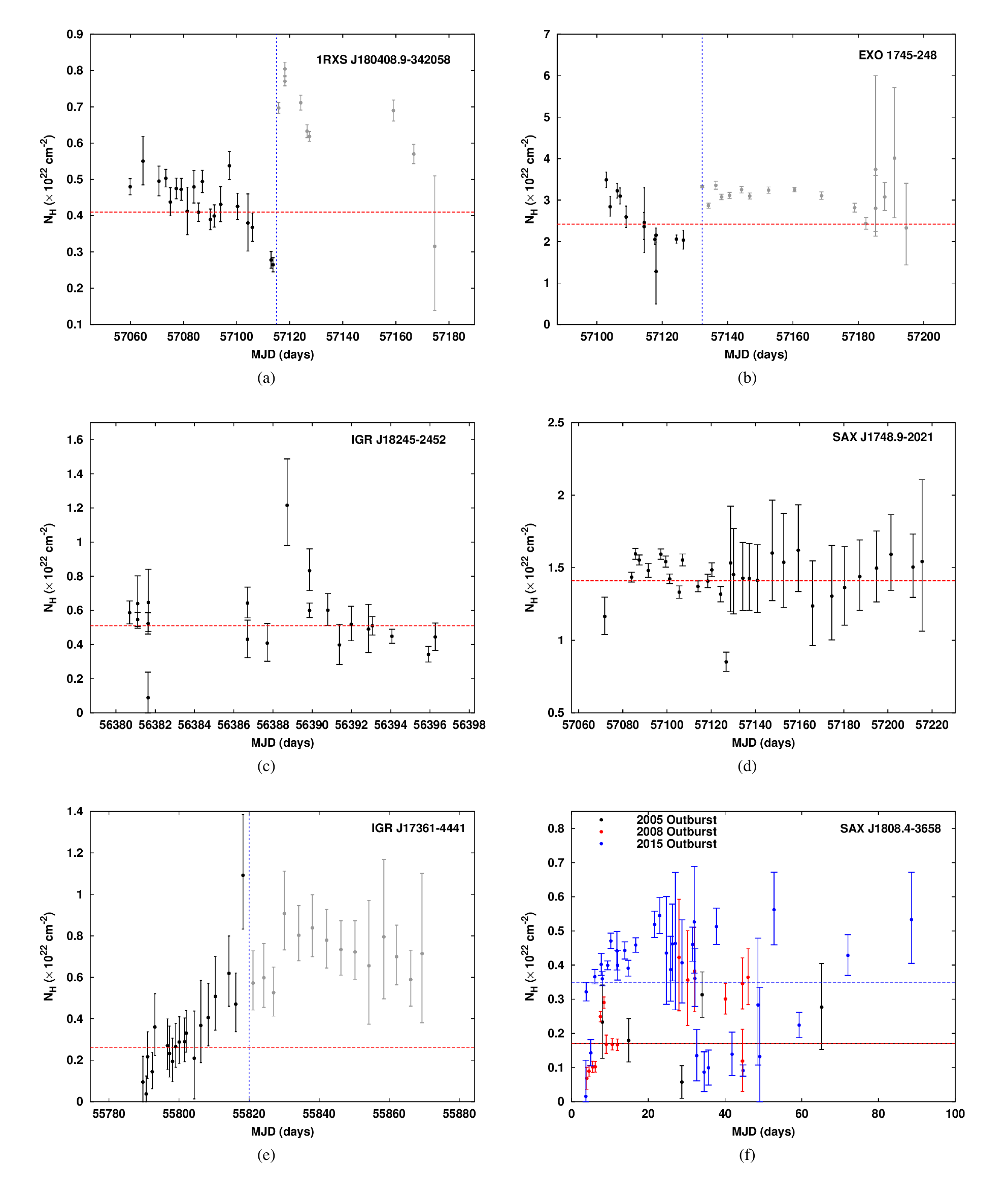}
\caption{The free $N_\text{H}$ evolution with time for our six sources. The best-fit $N_\text{H}$ is calculated from a constant fit through the relevant points for each source and shown by the horizontal dashed line (see Table 1 in the main paper for the best-fit $N_\text{H}$ values). We only consider the hard state data for the best-fit $N_\text{H}$ determination. A hard to soft transition has been reported in literature only for 1RXS J180408.9$-$342058 \citep{degenaar2016disk} and EXO 1745$-$248 \citep{tetarenko2016disc}, as shown by the vertical dashed blue line. The spectral information presented in \citet{del2014puzzling} hints at a state change for IGR J17361$-$4441 (indicated by the dashed blue line) therefore we only use the assumed hard state data. Thus for those sources the soft state $N_\text{H}$ values (shown in grey in (a), (b), and (e)) are not taken into account for the best-fit calculation. For SAX J1808.4$-$3658 (f) the best-fit $N_\text{H}$ values have been determined individually for the 3 different outbursts considered here. The zero point for the time axis in (f) are :  MJD = 53530 (2005 outburst), MJD = 54730 (2008 outburst), and MJD = 57120 (2015 outburst). The 2011 outburst was not considered due to poor coverage.}
\label{fig_nH_var}
\end{figure}

Here we discuss how we have calculated the best-fit $N_\text{H}$. The $N_\text{H}$ was left free (see Figure \ref{fig_nH_var}) and a constant was fit through the appropriate $N_\text{H}$ data to determine the best-fit $N_\text{H}$. We prefer to use hard state data to determine the best-fit $N_\text{H}$. To determine the different source states, we compared the {\it MAXI} data\footnote{http://maxi.riken.jp/top/slist.html} of each of the sources with that obtained using {\it Swift}/BAT\footnote{https://swift.gsfc.nasa.gov/results/transients/}. Unfortunately, only 1RXS J180408.9$-$342058 and EXO 1745$-$248 showed clearly different spectral states in those light curves \citep[see also][]{degenaar2016disk,tetarenko2016disc}. For 1RXS J180408.9$-$342058, the state transition was also accompanied by a change in the $N_\text{H}$ trend (see Figure \ref{fig_nH_var}). However, this is not so clearly seen for EXO 1745$-$248, possibly due to the intrinsic higher $N_\text{H}$ compared to 1RXS J180408.9$-$342058.

For the other sources, it could not be determined in which state the sources were at which moment. However, for IGR J18245$-$2452 and SAX J1748.9$-$2021 the $N_\text{H}$ does not seem to evolve as the outburst progresses. Therefore, for those two sources we used all the data points to determine the best-fit $N_\text{H}$. 

For IGR J17361$-$441 \citet{del2014puzzling} have shown that the data could be fitted with a two component model that included a soft component in addition to the power-law component. Using this model, they found that the soft component remained at approximately the same temperature throughout the full outburst, even when the fluxes decreased significantly.  This indicates that the spectra became softer when the luminosity decreased at the end of the outburst. Therefore,  we only used the data from when the flux remained constant (i.e. before MJD $\sim$ 55820, see Figure 1 of \citeauthor{del2014puzzling} \citeyear{del2014puzzling}) to calculate the best-fit $N_\text{H}$. We note that the best-fit $N_\text{H}$ for IGR J17361$-$4441 is only slightly higher than the Galactic $N_\text{H}$.  

SAX J1808.4$-$3658 has been observed over multiple outbursts. For the 2008 outburst, the spectrum near the peak $L_\text{X}$ was strongly dominated by a power-law component when fitted with a multiple component model (Patruno et al., 2009), suggesting that the source never transitioned to a soft state during this outburst. Although no detailed spectral study for its 2015 outburst has been published so far the $N_\text{H}$ evolution (Figure 1(f)) indicates that the source was significantly softer during the peak of the 2015 outburst than during the 2008 outburst because the $N_\text{H}$ was much higher (a factor of $\sim$2.5). However, all three outbursts studied here also show an $N_\text{H}$ evolution during each given outburst. Due to the different $N_\text{H}$ behaviour in the different outbursts it is difficult to determine how to calculate the best-fit $N_\text{H}$. The best-fit $N_\text{H}$ determined from only the 2008 outburst is 0.17 $\times 10^{22}$ cm$^{-2}$, which is the same value as determined from the 2005 outburst (and very close to the Galactic $N_\text{H}$ value of 0.14$\times 10^{22}$ cm$^{-2}$). The best-fit $N_\text{H}$ as determined from the 2015 outburst is 0.4$\times 10^{22}$ cm$^{-2}$. The best-fit $N_\text{H}$ as determined from all the outbursts is 0.35$\times 10^{22}$ cm$^{-2}$. Thus, it is not straightforward to select a best-fit $N_\text{H}$ for this source and, thus, we only use the $N_\text{H}$ left free and Galactic $N_\text{H}$ values to study the spectral evolution of the source.

\FloatBarrier

\section{Photon Index Evolution with Time}
\setcounter{figure}{0}

\begin{figure}
\centering
\includegraphics[scale=0.9]{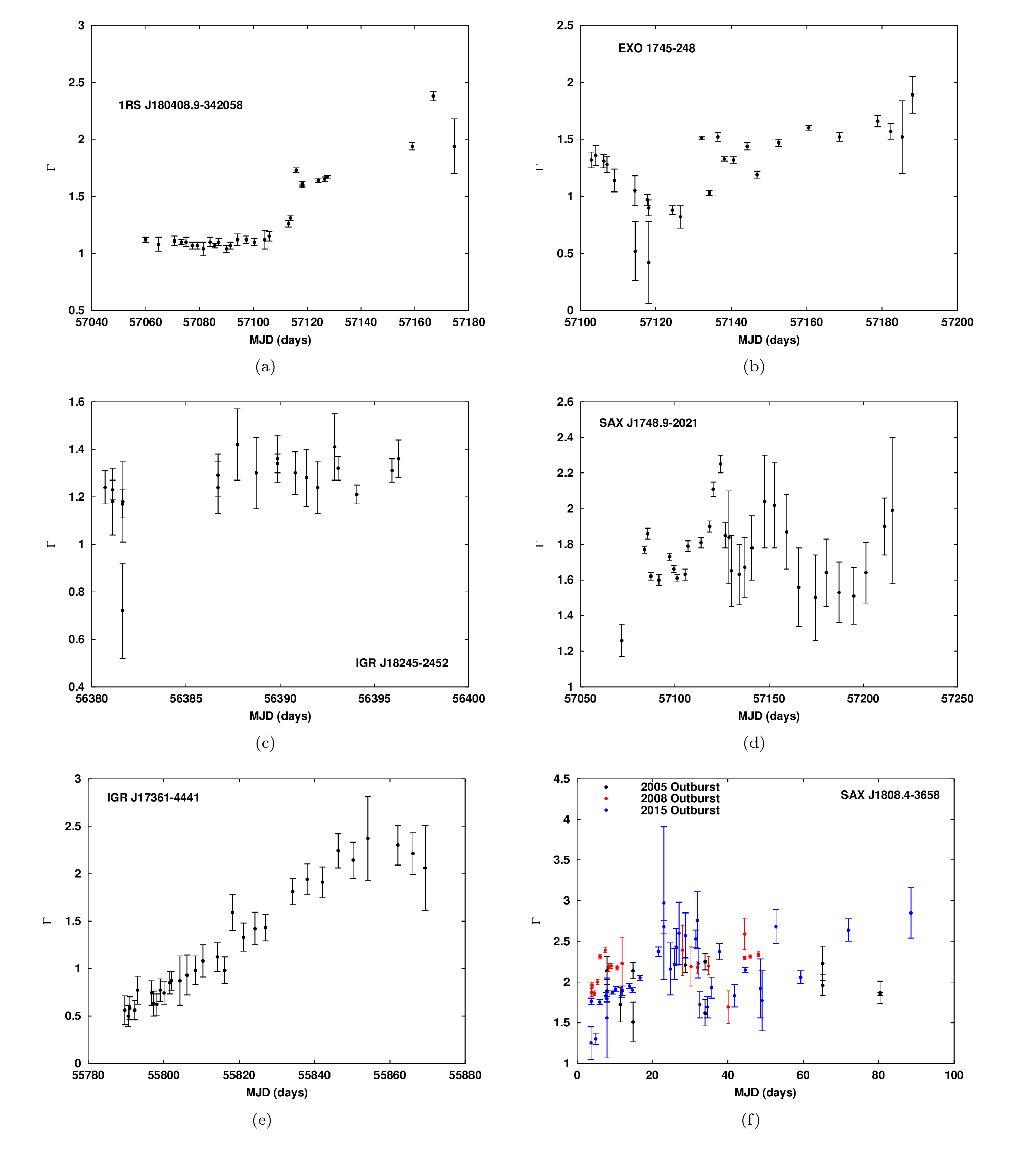}
\caption{The photon index $\Gamma$ evolution with time for our six sources when the $N_\mathrm{H}$ is left free. The values are tabulated in Appendix \ref{sect_tab_val}. The zero point for the time axis in (f) are :  MJD = 53530 (2005 outburst), MJD = 54730 (2008 outburst), and MJD = 57120 (2015 outburst). The 2011 outburst was not considered due to poor coverage.}
\label{fig_gamma_var}
\end{figure}

\FloatBarrier
\begin{figure}
\section{Individual Photon Index versus Luminosity plots for all the sources}
\label{sect_tab_val}
\setcounter{figure}{0}

\centering
\resizebox{17cm}{20cm}{%
\begin{tikzpicture}
\node[inner sep=0pt] (russell) at (0,0)
    {\includegraphics[scale=0.85]{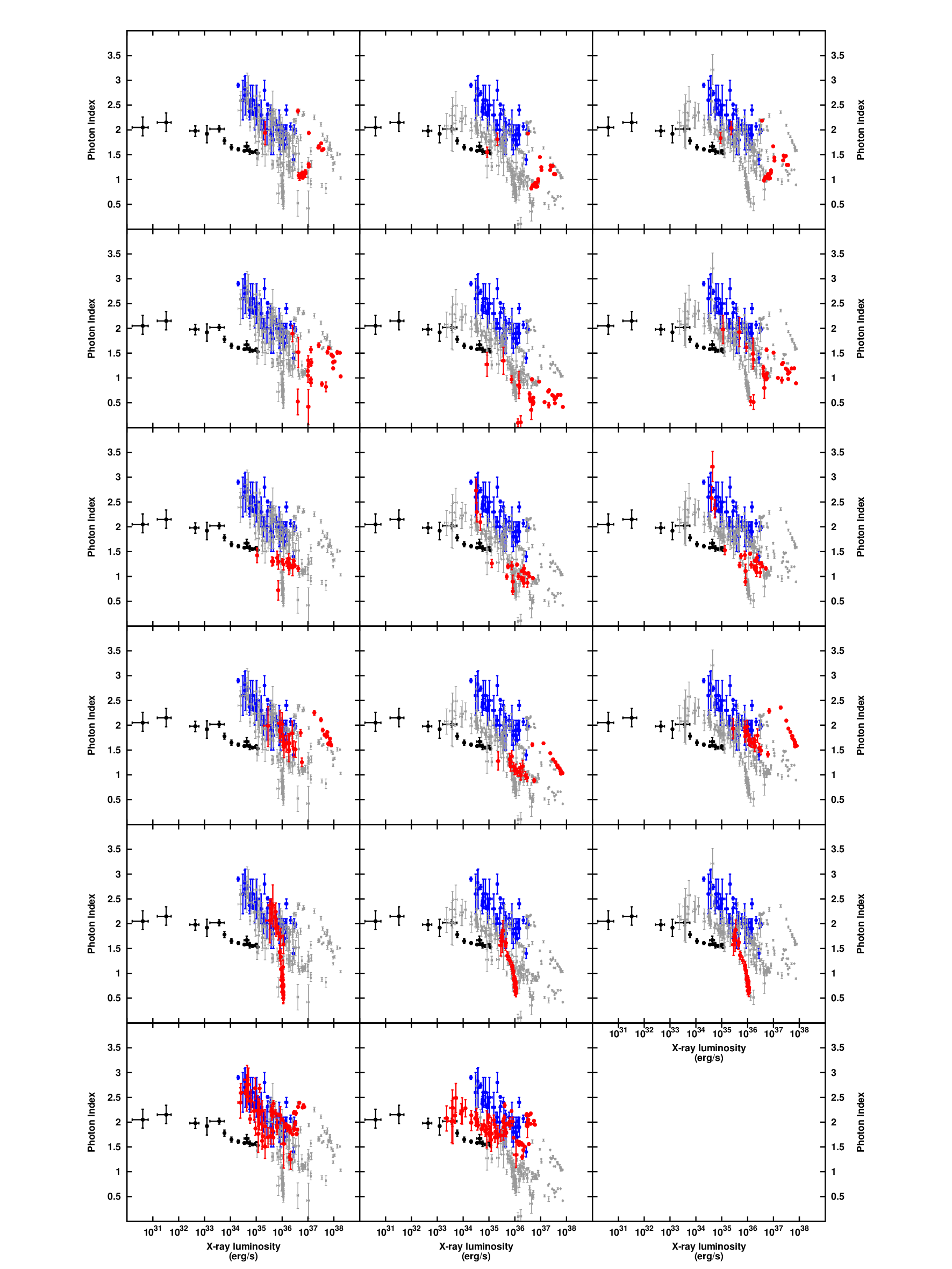}};
\fill[white] (9.5,-16) rectangle (12,16);
\fill[white] (3.5,-17) rectangle (9.5,-12);
\fill[white] (3.15,-16) rectangle (4,-15.5);
\fill[white] (9,-12) rectangle (9.6,-10.45);

% The legend
\node at (-7.4,11.4) [draw=none]  {\footnotesize{1RXS J180408.9-342058}};
\node at (-1.2,11.4) [draw=none]  {\footnotesize{1RXS J180408.9-342058}};
\node at (5,11.4) [draw=none]  {\footnotesize{1RXS J180408.9-342058}};

\node at (-7.8,6) [draw=none]  {\footnotesize{EXO 1745-248}};
\node at (-1.5,6) [draw=none]  {\footnotesize{EXO 1745-248}};
\node at (4.6,6) [draw=none]  {\footnotesize{EXO 1745-248}};

\node at (-7.8,0.7) [draw=none]  {\footnotesize{IGR J18245-2452}};
\node at (-1.5,0.7) [draw=none]  {\footnotesize{IGR J18245-2452}};
\node at (4.9,0.7) [draw=none]  {\footnotesize{IGR J18245-2452}};

\node at (-7.8,-4.6) [draw=none]  {\footnotesize{SAX J1748.9-2021}};
\node at (-1.6,-4.6) [draw=none]  {\footnotesize{SAX J1748.9-2021}};
\node at (4.6,-4.6) [draw=none]  {\footnotesize{SAX J1748.9-2021}};

\node at (-7.8,-10) [draw=none]  {\footnotesize{IGR J17361-4441}};
\node at (-1.5,-10) [draw=none]  {\footnotesize{IGR J17361-4441}};
\node at (4.8,-10) [draw=none]  {\footnotesize{IGR J17361-4441}};

\node at (-7.8,-15.25) [draw=none]  {\footnotesize{SAX J1808.4-3658}};
\node at (-1.6,-15.25) [draw=none]  {\footnotesize{SAX J1808.4-3658}};
%\node at (4.6,-15.25) [draw=none]  {\footnotesize{SAX J1808.4-3658}};

\node at (-6.3,17) [draw=none]  {\LARGE{$N_{\text{H}}$ left free}};
\node at (0,17) [draw=none]  {\LARGE{Galactic $N_{\text{H}}$}};
\node at (6.5,17) [draw=none]  {\LARGE{Best-fit $N_{\text{H}}$}};

\end{tikzpicture}
}
\caption{The photon index versus luminosity (0.5 -- 10 keV) for our six sources plotted for the $N_\text{H}$ left free, the Galactic $N_\text{H}$, and the best-fit $N_\text{H}$ value. The black data points and blue data points represent the black hole and neutron star data from \citet{wijnands2015low}. The red data points in each figure are the $\Gamma$ versus $L_X$ values for the individual source (the specific source name is give in the bottom left corner of each panel). The grey points are the data for the other sources in our sample. The photon index versus luminosity values for SAX J1808.4$-$3658 using the best-fit $N_\text{H}$ were not calculated (see Appendix \ref{sect_best_fit_app}) and therefore are not presented here.}
\end{figure}

\FloatBarrier
\clearpage
\section{Photon Index versus Luminosity}
In this appendix we tabulate the photon index ($\Gamma$) versus luminosity ($L_X$) values for the six sources in our sample. The values leaving $N_\text{H}$ free (first set of $N_\text{H}$ columns in the tables) as well as using the Galactic $N_\text{H}$ (second set) and the best-fit $N_\text{H}$ (third set) have been reported. If two rows are reported for the same observation ID it means that for these observations the data from both, the WT and PC mode, satisfied our selection criteria and were included in our result.

\setcounter{table}{0}

\begin{sidewaystable}

\centering
\caption{1RXS J180408.9$-$342058}

\begin{tabular}{p{2.1cm}p{1.0cm}p{1.5cm}P{5.9cm}P{3.5cm}P{3.5cm}}
\tabularnewline
\hline
&&&$N_\text{H}$ free&  $N_\text{H}$ =  0.20 $\times10^{22}$ cm$^{-2}$&  $N_\text{H}$ =  0.41 $\times10^{22}$ cm$^{-2}$\tabularnewline
\cmidrule(r){4-4}\cmidrule(lr){5-5}\cmidrule(l){6-6}
%\cmidrule{4-6}
\end{tabular}

\begin{tabular}{cccccccccc}
MJD& Obs ID & Exposure& $N_\text{H}$ & $\Gamma$ & L$_X$  &  $\Gamma$ & L$_X$&  $\Gamma$ & L$_X$\tabularnewline
(days)& & Time (ks)&(10$^{22}$ cm$^{-2}$)&&(10$^{35}$ erg s$^{-1}$)&&(10$^{35}$ erg s$^{-1}$)& &(10$^{35}$ erg s$^{-1}$)\tabularnewline
\hline

57059.88 	&	00630047000	&	3.67&	0.48 	 $\pm$  	0.02 	 &	1.12 	 $\pm$  	0.02 	 &	55.75 	 $\pm$  	0.50 	 &	0.90 	 $\pm$  	0.01 	 &	55.48 	 $\pm$  	0.52& 1.07 	 $\pm$  	0.01 	 &	55.55 	 $\pm$  	0.50\tabularnewline  
57064.73 	&	00032436009	&	0.83&	0.55 	 $\pm$  	0.07 	 &	1.08 	 $\pm$  	0.06 	 &	43.52 	 $\pm$  	1.06 	 &	0.82 	 $\pm$  	0.03 	 &	43.39 	 $\pm$  	1.14& 0.98 	 $\pm$  	0.03 	 &	43.28 	 $\pm$  	1.08\tabularnewline  
57070.73 	&	00032436010	&	0.96&	0.50 	 $\pm$  	0.04 	 &	1.11 	 $\pm$  	0.04 	 &	63.26 	 $\pm$  	1.04 	 &	0.88 	 $\pm$  	0.02 	 &	63.07 	 $\pm$  	1.11& 1.05 	 $\pm$  	0.02 	 &	63.03 	 $\pm$  	1.05\tabularnewline  
57073.32 	&	00032436014	&	2.76&	0.50 	 $\pm$  	0.02 	 &	1.10 	 $\pm$  	0.02 	 &	62.60 	 $\pm$  	0.60 	 &	0.87 	 $\pm$  	0.01 	 &	62.39 	 $\pm$  	0.64& 1.03 	 $\pm$  	0.01 	 &	62.35 	 $\pm$  	0.61\tabularnewline  
57075.04 	&	00032436015	&	0.97&	0.44 	 $\pm$  	0.04 	 &	1.10 	 $\pm$  	0.04 	 &	64.70 	 $\pm$  	1.05 	 &	0.91 	 $\pm$  	0.02 	 &	64.51 	 $\pm$  	1.11& 1.08 	 $\pm$  	0.02 	 &	64.62 	 $\pm$  	1.05\tabularnewline  
57077.25 	&	00032436016	&	1.96&	0.47 	 $\pm$  	0.03 	 &	1.07 	 $\pm$  	0.03 	 &	61.44 	 $\pm$  	0.70 	 &	0.86 	 $\pm$  	0.01 	 &	61.34 	 $\pm$  	0.75& 1.02 	 $\pm$  	0.02 	 &	61.29 	 $\pm$  	0.71\tabularnewline  
57079.11 	&	00032436017	&	1.72&	0.47 	 $\pm$  	0.03 	 &	1.07 	 $\pm$  	0.03 	 &	68.55 	 $\pm$  	0.84 	 &	0.86 	 $\pm$  	0.02 	 &	68.46 	 $\pm$  	0.89& 1.02 	 $\pm$  	0.02 	 &	68.39 	 $\pm$  	0.84\tabularnewline  
57081.38 	&	00032436018	&	0.55&	0.41 	 $\pm$  	0.07 	 &	1.04 	 $\pm$  	0.06 	 &	47.65 	 $\pm$  	1.30 	 &	0.87 	 $\pm$  	0.03 	 &	47.64 	 $\pm$  	1.37& 1.04 	 $\pm$  	0.04 	 &	47.65 	 $\pm$  	1.30\tabularnewline  
57083.90 	&	00032436019	&	1.08&	0.48 	 $\pm$  	0.05 	 &	1.10 	 $\pm$  	0.04 	 &	58.62 	 $\pm$  	1.05 	 &	0.88 	 $\pm$  	0.02 	 &	58.42 	 $\pm$  	1.12& 1.05 	 $\pm$  	0.02 	 &	58.44 	 $\pm$  	1.06\tabularnewline  
57085.63 	&	00032436020	&	2.15&	0.41 	 $\pm$  	0.03 	 &	1.07 	 $\pm$  	0.02 	 &	64.01 	 $\pm$  	0.69 	 &	0.90 	 $\pm$  	0.01 	 &	63.93 	 $\pm$  	0.72& 1.07 	 $\pm$  	0.01 	 &	64.01 	 $\pm$  	0.68\tabularnewline  
57087.09 	&	00081436001	&	1.69&	0.49 	 $\pm$  	0.03 	 &	1.10 	 $\pm$  	0.03 	 &	66.39 	 $\pm$  	0.81 	 &	0.87 	 $\pm$  	0.02 	 &	66.22 	 $\pm$  	0.86& 1.03 	 $\pm$  	0.02 	 &	66.15 	 $\pm$  	0.82\tabularnewline  
57090.09 	&	00032436021	&	1.71&	0.39 	 $\pm$  	0.03 	 &	1.04 	 $\pm$  	0.03 	 &	62.51 	 $\pm$  	0.77 	 &	0.88 	 $\pm$  	0.02 	 &	62.52 	 $\pm$  	0.81& 1.05 	 $\pm$  	0.02 	 &	62.55 	 $\pm$  	0.77\tabularnewline  
57091.56 	&	00032436022	&	1.49&	0.40 	 $\pm$  	0.03 	 &	1.07 	 $\pm$  	0.03 	 &	62.05 	 $\pm$  	0.82 	 &	0.91 	 $\pm$  	0.02 	 &	62.02 	 $\pm$  	0.86& 1.08 	 $\pm$  	0.02 	 &	62.08 	 $\pm$  	0.82\tabularnewline  
57094.01 	&	00032436023	&	0.59&	0.43 	 $\pm$  	0.05 	 &	1.12 	 $\pm$  	0.05 	 &	61.97 	 $\pm$  	1.27 	 &	0.94 	 $\pm$  	0.03 	 &	61.73 	 $\pm$  	1.33& 1.11 	 $\pm$  	0.03 	 &	61.90 	 $\pm$  	1.26\tabularnewline  
57097.28 	&	00032436024	&	0.97&	0.54 	 $\pm$  	0.04 	 &	1.12 	 $\pm$  	0.03 	 &	74.20 	 $\pm$  	1.10 	 &	0.86 	 $\pm$  	0.02 	 &	73.84 	 $\pm$  	1.17& 1.03 	 $\pm$  	0.02 	 &	73.76 	 $\pm$  	1.11\tabularnewline  
57100.35 	&	00032436025	&	0.97&	0.43 	 $\pm$  	0.04 	 &	1.10 	 $\pm$  	0.03 	 &	67.74 	 $\pm$  	1.04 	 &	0.92 	 $\pm$  	0.02 	 &	67.55 	 $\pm$  	1.09& 1.09 	 $\pm$  	0.02 	 &	67.69 	 $\pm$  	1.03\tabularnewline  
57104.26 	&	00032436026	&	0.18&	0.38 	 $\pm$  	0.08 	 &	1.12 	 $\pm$  	0.08 	 &	80.19 	 $\pm$  	2.79 	 &	0.97 	 $\pm$  	0.04 	 &	79.96 	 $\pm$  	2.92& 1.14 	 $\pm$  	0.05 	 &	80.32 	 $\pm$  	2.75\tabularnewline  
57105.94 	&	00032436027	&	0.58&	0.37 	 $\pm$  	0.04 	 &	1.15 	 $\pm$  	0.04 	 &	79.88 	 $\pm$  	1.43 	 &	1.01 	 $\pm$  	0.02 	 &	79.59 	 $\pm$  	1.49& 1.18 	 $\pm$  	0.02 	 &	80.09 	 $\pm$  	1.41\tabularnewline  
57112.98 	&	00032436028	&	0.97&	0.28 	 $\pm$  	0.02 	 &	1.26 	 $\pm$  	0.03 	 &	107.54 	 $\pm$  	1.23 	 &	1.19 	 $\pm$  	0.01 	 &	107.02 	 $\pm$  	1.24& 1.38 	 $\pm$  	0.02 	 &	109.15 	 $\pm$  	1.17\tabularnewline  
57113.71 	&	00081436002	&	1.84&	0.26 	 $\pm$  	0.02 	 &	1.31 	 $\pm$  	0.02 	 &	105.91 	 $\pm$  	1.06 	 &	1.25 	 $\pm$  	0.01 	 &	105.35 	 $\pm$  	1.07& 1.44 	 $\pm$  	0.01 	 &	108.00 	 $\pm$  	1.00\tabularnewline  
57115.84 	&	00032436029	&	1.15&	0.70 	 $\pm$  	0.01 	 &	1.73 	 $\pm$  	0.02 	 &	322.11 	 $\pm$  	2.26 	 &	1.28 	 $\pm$  	0.01 	 &	291.92 	 $\pm$  	1.85& 1.48 	 $\pm$  	0.01 	 &	299.77 	 $\pm$  	1.74\tabularnewline  
57118.18 	&	00032436030	&	0.97&	0.80 	 $\pm$  	0.02 	 &	1.61 	 $\pm$  	0.02 	 &	397.97 	 $\pm$  	2.96 	 &	1.11 	 $\pm$  	0.01 	 &	365.62 	 $\pm$  	2.62& 1.29 	 $\pm$  	0.01 	 &	370.26 	 $\pm$  	2.46\tabularnewline  
57118.18 	&	00032436031	&	1.93&	0.77 	 $\pm$  	0.01 	 &	1.59 	 $\pm$  	0.01 	 &	349.92 	 $\pm$  	1.78 	 &	1.11 	 $\pm$  	0.01 	 &	324.08 	 $\pm$  	1.59& 1.30 	 $\pm$  	0.01 	 &	328.26 	 $\pm$  	1.51\tabularnewline  
57124.17 	&	00032436032	&	0.96&	0.71 	 $\pm$  	0.02 	 &	1.64 	 $\pm$  	0.02 	 &	250.86 	 $\pm$  	2.23 	 &	1.19 	 $\pm$  	0.01 	 &	231.23 	 $\pm$  	1.96& 1.38 	 $\pm$  	0.01 	 &	235.67 	 $\pm$  	1.85\tabularnewline  
57126.54 	&	00081451001	&	1.00&	0.63 	 $\pm$  	0.02 	 &	1.65 	 $\pm$  	0.02 	 &	277.20 	 $\pm$  	2.26 	 &	1.26 	 $\pm$  	0.01 	 &	257.05 	 $\pm$  	1.98& 1.46 	 $\pm$  	0.01 	 &	263.59 	 $\pm$  	1.87\tabularnewline  
57127.41 	&	00081451002	&	1.81&	0.62 	 $\pm$  	0.01 	 &	1.67 	 $\pm$  	0.01 	 &	259.90 	 $\pm$  	1.68 	 &	1.28 	 $\pm$  	0.01 	 &	240.77 	 $\pm$  	1.46& 1.48 	 $\pm$  	0.01 	 &	247.42 	 $\pm$  	1.37\tabularnewline  
57159.03 	&	00032436033	&	0.74&	0.69 	 $\pm$  	0.03 	 &	1.94 	 $\pm$  	0.03 	 &	107.74 	 $\pm$  	1.78 	 &	1.45 	 $\pm$  	0.02 	 &	93.04 	 $\pm$  	1.17& 1.67 	 $\pm$  	0.02 	 &	97.30 	 $\pm$  	1.11\tabularnewline  
57166.76 	&	00032436034	&	1.62&	0.57 	 $\pm$  	0.03 	 &	2.38 	 $\pm$  	0.04 	 &	40.58 	 $\pm$  	0.98 	 &	1.93 	 $\pm$  	0.02 	 &	32.11 	 $\pm$  	0.38& 2.19 	 $\pm$  	0.02 	 &	36.12 	 $\pm$  	0.38\tabularnewline  
57174.65 	&	00032436035	&	0.93&	0.32 	 $\pm$  	0.19 	 &	1.94 	 $\pm$  	0.24 	 &	2.20 	 $\pm$  	0.27 	 &	1.81 	 $\pm$  	0.13 	 &	2.10 	 $\pm$  	0.17& 2.04 	 $\pm$  	0.13 	 &	2.30 	 $\pm$  	0.17\tabularnewline  
57175.58 	&	00033806001	&	1.98&	- 	 &	- 	 &	1.05 	 $\pm$  	0.14 	 &	1.55 	 $\pm$  	0.10 	 &	0.85 	 $\pm$  	0.07& 1.84 	 $\pm$  	0.11 	 &	0.90 	 $\pm$  	0.06\tabularnewline

\hline
\end{tabular}
\end{sidewaystable}

\begin{sidewaystable}
\small
\centering
\caption{EXO 1745$-$248}

\begin{tabular}{p{1.8cm}p{0.8cm}p{1.3cm}P{5.5cm}P{3.4cm}P{3.4cm}}
\tabularnewline
\hline
&&&$N_\text{H}$ free&  $N_\text{H}$ =  1.10 $\times10^{22}$ cm$^{-2}$&  $N_\text{H}$ =  2.42 $\times10^{22}$ cm$^{-2}$\tabularnewline
\cmidrule(r){4-4}\cmidrule(lr){5-5}\cmidrule(l){6-6}
\end{tabular}

\begin{tabular}{cccccccccc}
MJD& Obs ID & Exposure& $N_\text{H}$ & $\Gamma$ & L$_X$  &  $\Gamma$ & L$_X$&  $\Gamma$ & L$_X$\tabularnewline
(days)& & Time (ks)&(10$^{22}$ cm$^{-2}$)&&(10$^{35}$ erg s$^{-1}$)&&(10$^{35}$ erg s$^{-1}$)& &(10$^{35}$ erg s$^{-1}$)\tabularnewline
\hline

57098.74 	&	00032148017	&	1.97&	-		 	 &	-			 &	-		     	 &	0.10 	 $\pm$  	0.08 	 &	13.20 	 $\pm$  	0.69  &	0.53 	 $\pm$  	0.09 	 &	13.17 	 $\pm$  	0.63\tabularnewline  
57102.87 	&	00032148021	&	1.19	&	3.49 	 $\pm$  	0.19 	 &	1.32 	 $\pm$  	0.07 	 &	57.13 	 $\pm$  	1.58 	 &	0.47 	 $\pm$  	0.03 	 &	49.41 	 $\pm$  	0.98  &	0.98 	 $\pm$  	0.03 	 &	51.67 	 $\pm$  	0.86\tabularnewline  
57104.06 	&	00032148023	&	0.77	&	2.84 	 $\pm$  	0.25 	 &	1.36 	 $\pm$  	0.09 	 &	42.70 	 $\pm$  	1.73 	 &	0.68 	 $\pm$  	0.04 	 &	37.21 	 $\pm$  	1.02  &	1.21 	 $\pm$  	0.05 	 &	40.60 	 $\pm$  	0.90\tabularnewline  
57106.20 	&	00032148024	&	1.27	&	3.22 	 $\pm$  	0.18 	 &	1.31 	 $\pm$  	0.06 	 &	55.67 	 $\pm$  	1.48 	 &	0.54 	 $\pm$  	0.03 	 &	48.60 	 $\pm$  	0.94  &	1.05 	 $\pm$  	0.03 	 &	51.43 	 $\pm$  	0.82\tabularnewline  
57107.07 	&	00032148025	&	1.30	&	3.10 	 $\pm$  	0.20 	 &	1.28 	 $\pm$  	0.07 	 &	45.04 	 $\pm$  	1.28 	 &	0.55 	 $\pm$  	0.03 	 &	39.86 	 $\pm$  	0.84  &	1.06 	 $\pm$  	0.04 	 &	42.16 	 $\pm$  	0.73\tabularnewline  
57108.94 	&	00032148026	&	0.64	&	2.59 	 $\pm$  	0.26 	 &	1.14 	 $\pm$  	0.10 	 &	41.51 	 $\pm$  	1.50 	 &	0.58 	 $\pm$  	0.05 	 &	38.18 	 $\pm$  	1.20  &	1.08 	 $\pm$  	0.05 	 &	40.86 	 $\pm$  	1.05\tabularnewline  
57114.44 	&	00032148027	&	0.35	&	2.36 	 $\pm$  	0.34 	 &	1.05 	 $\pm$  	0.13 	 &	42.11 	 $\pm$  	1.92 	 &	0.57 	 $\pm$  	0.06 	 &	39.66 	 $\pm$  	1.68  &	1.07 	 $\pm$  	0.07 	 &	42.32 	 $\pm$  	1.48\tabularnewline  
57114.51 	&	00032148027	&	0.68	&	2.46 	 $\pm$  	0.84 	 &	0.52 	 $\pm$  	0.26 	 &	16.98 	 $\pm$  	1.36 	 &	0.11 	 $\pm$  	0.13 	 &	17.02 	 $\pm$  	1.49  &	0.51 	 $\pm$  	0.14 	 &	16.97 	 $\pm$  	1.35\tabularnewline  
57117.77 	&	00637212000	&	2.08	&	2.05 	 $\pm$  	0.12 	 &	0.97 	 $\pm$  	0.05 	 &	53.95 	 $\pm$  	0.81 	 &	0.61 	 $\pm$  	0.02 	 &	51.95 	 $\pm$  	0.75  &	1.09 	 $\pm$  	0.03 	 &	55.48 	 $\pm$  	0.67\tabularnewline  
57118.11 	&	00032148029	&	0.86	&	2.15 	 $\pm$  	0.17 	 &	0.90 	 $\pm$  	0.07 	 &	54.50 	 $\pm$  	1.20 	 &	0.51 	 $\pm$  	0.03 	 &	52.70 	 $\pm$  	1.19  &	0.99 	 $\pm$  	0.04 	 &	55.44 	 $\pm$  	1.05\tabularnewline  
57118.12 	&	00032148029	&	0.13	&	1.28 	 $\pm$  	0.89 	 &	0.42 	 $\pm$  	0.36 	 &	43.49 	 $\pm$  	5.80 	 &	0.36 	 $\pm$  	0.19 	 &	43.56 	 $\pm$  	5.85  &	0.80 	 $\pm$  	0.21 	 &	44.54 	 $\pm$  	5.20\tabularnewline  
57124.35 	&	00032148030	&	0.83	&	2.06 	 $\pm$  	0.10 	 &	0.88 	 $\pm$  	0.04 	 &	145.09 	 $\pm$  	1.93 	 &	0.51 	 $\pm$  	0.02 	 &	141.06 	 $\pm$  	1.92  &	1.00 	 $\pm$  	0.02 	 &	148.48 	 $\pm$  	1.72\tabularnewline  
57126.48 	&	00032148032	&	0.85	&	2.04 	 $\pm$  	0.23 	 &	0.82 	 $\pm$  	0.10 	 &	206.97 	 $\pm$  	6.78 	 &	0.45 	 $\pm$  	0.05 	 &	204.29 	 $\pm$  	7.20  &	0.96 	 $\pm$  	0.06 	 &	211.17 	 $\pm$  	6.31\tabularnewline  
57132.22 	&	00032148033	&	3.18	&	3.33 	 $\pm$  	0.03 	 &	1.51 	 $\pm$  	0.01 	 &	718.57 	 $\pm$  	3.87 	 &	0.66 	 $\pm$  	0.00 	 &	586.74 	 $\pm$  	1.94  &	1.20 	 $\pm$  	0.01 	 &	635.80 	 $\pm$  	1.70\tabularnewline  
57134.14 	&	00032148034	&	0.68	&	2.87 	 $\pm$  	0.06 	 &	1.03 	 $\pm$  	0.02 	 &	769.00 	 $\pm$  	5.60 	 &	0.42 	 $\pm$  	0.01 	 &	719.21 	 $\pm$  	4.95  &	0.89 	 $\pm$  	0.01 	 &	746.01 	 $\pm$  	4.36\tabularnewline  
57136.40 	&	00032148035	&	0.38	&	3.36 	 $\pm$  	0.10 	 &	1.52 	 $\pm$  	0.04 	 &	569.66 	 $\pm$  10.39   	 &	0.66 	 $\pm$  	0.02 	 &	462.04 	 $\pm$  	4.97  &	1.19 	 $\pm$  	0.02 	 &	501.57 	 $\pm$  	4.37\tabularnewline  
57138.14 	&	00032148036	&	0.99	&	3.08 	 $\pm$  	0.07 	 &	1.33 	 $\pm$  	0.02 	 &	449.23 	 $\pm$  	4.54 	 &	0.60 	 $\pm$  	0.01 	 &	390.06 	 $\pm$  	2.84  &	1.11 	 $\pm$  	0.01 	 &	417.98 	 $\pm$  	2.49\tabularnewline  
57140.59 	&	00032148037	&	0.96	&	3.12 	 $\pm$  	0.07 	 &	1.32 	 $\pm$  	0.03 	 &	392.91 	 $\pm$  	4.30 	 &	0.57 	 $\pm$  	0.01 	 &	343.29 	 $\pm$  	2.76  &	1.09 	 $\pm$  	0.01 	 &	365.28 	 $\pm$  	2.42\tabularnewline  
57144.25 	&	00032148038	&	0.84	&	3.25 	 $\pm$  	0.08 	 &	1.44 	 $\pm$  	0.03 	 &	383.66 	 $\pm$  	5.27 	 &	0.63 	 $\pm$  	0.01 	 &	321.40 	 $\pm$  	2.85  &	1.16 	 $\pm$  	0.02 	 &	346.46 	 $\pm$  	2.47\tabularnewline  
57146.79 	&	00032148039	&	1.03	&	3.10 	 $\pm$  	0.07 	 &	1.19 	 $\pm$  	0.03 	 &	394.28 	 $\pm$  	3.82 	 &	0.49 	 $\pm$  	0.01 	 &	354.25 	 $\pm$  	2.75  &	0.98 	 $\pm$  	0.01 	 &	372.00 	 $\pm$  	2.45\tabularnewline  
57152.57 	&	00032148041	&	0.99	&	3.24 	 $\pm$  	0.08 	 &	1.47 	 $\pm$  	0.03 	 &	335.37 	 $\pm$  	4.57 	 &	0.66 	 $\pm$  	0.01 	 &	278.25 	 $\pm$  	2.37  &	1.19 	 $\pm$  	0.01 	 &	301.58 	 $\pm$  	2.08\tabularnewline  
57160.48 	&	00032148043	&	2.94	&	3.25 	 $\pm$  	0.05 	 &	1.60 	 $\pm$  	0.02 	 &	266.36 	 $\pm$  	2.92 	 &	0.75 	 $\pm$  	0.01 	 &	210.61 	 $\pm$  	1.25  &	1.31 	 $\pm$  	0.01 	 &	233.76 	 $\pm$  	1.11\tabularnewline  
57168.74 	&	00032148045	&	0.89	&	3.11 	 $\pm$  	0.09 	 &	1.52 	 $\pm$  	0.04 	 &	236.93 	 $\pm$  	4.20 	 &	0.73 	 $\pm$  	0.02 	 &	194.48 	 $\pm$  	2.06  &	1.28 	 $\pm$  	0.02 	 &	214.47 	 $\pm$  	1.83\tabularnewline  
57178.85 	&	00032148047	&	1.17	&	2.82 	 $\pm$  	0.11 	 &	1.66 	 $\pm$  	0.05 	 &	108.85 	 $\pm$  	2.75 	 &	0.93 	 $\pm$  	0.02 	 &	86.29 	 $\pm$  	1.12  &	1.51 	 $\pm$  	0.02 	 &	100.85 	 $\pm$  	1.06\tabularnewline  
57182.36 	&	00032148048	&	1.05	&	2.43 	 $\pm$  	0.14 	 &	1.57 	 $\pm$  	0.07 	 &	54.31 	 $\pm$  	1.74 	 &	0.97 	 $\pm$  	0.03 	 &	45.65 	 $\pm$  	0.85  &	1.57 	 $\pm$  	0.03 	 &	54.16 	 $\pm$  	0.83\tabularnewline  
57185.31 	&	00032148050	&	0.06	&	2.80 	 $\pm$  	0.79 	 &	1.52 	 $\pm$  	0.32 	 &	17.71 	 $\pm$  	3.28 	 &	0.86 	 $\pm$  	0.27 	 &	14.40 	 $\pm$  	2.51  &	1.48 	 $\pm$  	0.30 	 &	16.07 	 $\pm$  	2.10\tabularnewline  
57185.31 	&	00032148050	&	0.42&	-		 	 &	-			 &	-		     	 &	0.82 	 $\pm$  	0.13 	 &	14.95 	 $\pm$  	1.33  &	1.38 	 $\pm$  	0.15 	 &	16.69 	 $\pm$  	1.18\tabularnewline  
57188.09 	&	00032148051	&	3.62	&	3.08 	 $\pm$  	0.35 	 &	1.89 	 $\pm$  	0.16 	 &	10.55 	 $\pm$  	1.23 	 &	0.97 	 $\pm$  	0.07 	 &	7.45 	 $\pm$  	0.35  &	1.62 	 $\pm$  	0.08 	 &	8.98 	 $\pm$  	0.34\tabularnewline  
57191.08 	&	00032148052	&	0.25&	-		 	 &	-			 &	-		     	 &	1.35 	 $\pm$  	0.27 	 &	3.64 	 $\pm$  	0.58  &	1.92 	 $\pm$  	0.30 	 &	4.83 	 $\pm$  	0.77\tabularnewline  
57194.68 	&	00032148053	&	1.11&	-		 	 &	-			 &	-		     	 &	1.27 	 $\pm$  	0.24 	 &	0.84 	 $\pm$  	0.13  &	1.98 	 $\pm$  	0.29 	 &	1.12 	 $\pm$  	0.16\tabularnewline

\hline
\end{tabular}
\end{sidewaystable}

\begin{sidewaystable}
\centering
\caption{IGR J18245$-$2452}

\begin{tabular}{p{2.1cm}p{1.0cm}p{1.5cm}P{5.9cm}P{3.5cm}P{3.5cm}}
\tabularnewline
\hline
&&&$N_\text{H}$ free&  $N_\text{H}$ =  0.26 $\times10^{22}$ cm$^{-2}$&  $N_\text{H}$ =  0.51 $\times10^{22}$ cm$^{-2}$\tabularnewline
\cmidrule(r){4-4}\cmidrule(lr){5-5}\cmidrule(l){6-6}
\end{tabular}

\begin{tabular}{cccccccccc}
MJD& Obs ID & Exposure& $N_\text{H}$ & $\Gamma$ & L$_X$  &  $\Gamma$ & L$_X$&  $\Gamma$ & L$_X$\tabularnewline
(days)& & Time (ks)&(10$^{22}$ cm$^{-2}$)&&(10$^{35}$ erg s$^{-1}$)&&(10$^{35}$ erg s$^{-1}$)& &(10$^{35}$ erg s$^{-1}$)\tabularnewline
\hline

56380.70 	&	00032785001	&	1.97&	0.59 	 $\pm$  	0.07 	 &	1.24 	 $\pm$  	0.07 	 &	21.49 	 $\pm$  	0.61 	 &	0.96 	 $\pm$  	0.04 	 &	21.30 	 $\pm$  	0.65&	1.18 	 $\pm$  	0.04 	 &	21.36 	 $\pm$  	0.61\tabularnewline  
56381.10 	&	00552336000	&	1.35&	0.55 	 $\pm$  	0.04 	 &	1.23 	 $\pm$  	0.04 	 &	37.18 	 $\pm$  	0.60 	 &	1.00 	 $\pm$  	0.02 	 &	36.66 	 $\pm$  	0.63&	1.20 	 $\pm$  	0.02 	 &	37.06 	 $\pm$  	0.59\tabularnewline  
56381.10 	&	00552336000	&	0.58&	0.64 	 $\pm$  	0.16 	 &	1.18 	 $\pm$  	0.14 	 &	22.17 	 $\pm$  	1.35 	 &	0.87 	 $\pm$  	0.08 	 &	22.09 	 $\pm$  	1.46&	1.08 	 $\pm$  	0.08 	 &	22.01 	 $\pm$  	1.37\tabularnewline  
56381.63 	&	00552369000	&	0.46&	0.52 	 $\pm$  	0.06 	 &	1.17 	 $\pm$  	0.06 	 &	50.61 	 $\pm$  	1.24 	 &	0.97 	 $\pm$  	0.03 	 &	50.11 	 $\pm$  	1.29&	1.16 	 $\pm$  	0.03 	 &	50.56 	 $\pm$  	1.22\tabularnewline  
56381.63 	&	00552369000	&	0.15&	0.09 	 $\pm$  	0.15 	 &	0.72 	 $\pm$  	0.20 	 &	8.56 	 $\pm$  	1.09 	 &	0.89 	 $\pm$  	0.13 	 &	8.39 	 $\pm$  	1.00&	1.11 	 $\pm$  	0.14 	 &	8.38 	 $\pm$  	0.93\tabularnewline  
56381.65 	&	00552370000	&	0.31&	0.65 	 $\pm$  	0.19 	 &	1.18 	 $\pm$  	0.17 	 &	30.92 	 $\pm$  	2.17 	 &	0.87 	 $\pm$  	0.09 	 &	30.80 	 $\pm$  	2.37&	1.08 	 $\pm$  	0.09 	 &	30.69 	 $\pm$  	2.21\tabularnewline  
56386.70 	&	00032785002	&	0.27&	0.43 	 $\pm$  	0.11 	 &	1.24 	 $\pm$  	0.11 	 &	22.35 	 $\pm$  	1.06 	 &	1.09 	 $\pm$  	0.06 	 &	22.15 	 $\pm$  	1.10&	1.30 	 $\pm$  	0.06 	 &	22.52 	 $\pm$  	1.03\tabularnewline  
56386.70 	&	00032785002	&	1.51&	0.64 	 $\pm$  	0.09 	 &	1.29 	 $\pm$  	0.09 	 &	18.43 	 $\pm$  	0.68 	 &	0.96 	 $\pm$  	0.05 	 &	18.20 	 $\pm$  	0.72&	1.18 	 $\pm$  	0.05 	 &	18.22 	 $\pm$  	0.67\tabularnewline  
56387.71 	&	00032785003	&	2.12&	0.41 	 $\pm$  	0.11 	 &	1.42 	 $\pm$  	0.15 	 &	1.30 	 $\pm$  	0.09 	 &	1.26 	 $\pm$  	0.08 	 &	1.29 	 $\pm$  	0.09&	1.53 	 $\pm$  	0.09 	 &	1.32 	 $\pm$  	0.08\tabularnewline  
56388.72 	&	00032785004	&	2.11&	1.22 	 $\pm$  	0.27 	 &	1.30 	 $\pm$  	0.15 	 &	8.68 	 $\pm$  	0.50 	 &	0.70 	 $\pm$  	0.07 	 &	8.32 	 $\pm$  	0.46&	0.89 	 $\pm$  	0.07 	 &	8.25 	 $\pm$  	0.43\tabularnewline  
56389.86 	&	00032785005	&	1.41&	0.60 	 $\pm$  	0.04 	 &	1.34 	 $\pm$  	0.04 	 &	34.39 	 $\pm$  	0.56 	 &	1.07 	 $\pm$  	0.02 	 &	33.49 	 $\pm$  	0.58&	1.27 	 $\pm$  	0.02 	 &	34.02 	 $\pm$  	0.54\tabularnewline  
56389.86 	&	00032785005	&	1.04&	0.83 	 $\pm$  	0.13 	 &	1.36 	 $\pm$  	0.10 	 &	21.89 	 $\pm$  	0.90 	 &	0.90 	 $\pm$  	0.05 	 &	21.24 	 $\pm$  	0.92&	1.12 	 $\pm$  	0.05 	 &	21.20 	 $\pm$  	0.85\tabularnewline  
56390.79 	&	00032785006	&	1.93&	0.60 	 $\pm$  	0.10 	 &	1.30 	 $\pm$  	0.09 	 &	4.93 	 $\pm$  	0.20 	 &	1.00 	 $\pm$  	0.05 	 &	4.87 	 $\pm$  	0.21&	1.23 	 $\pm$  	0.05 	 &	4.88 	 $\pm$  	0.19\tabularnewline  
56391.39 	&	00032785007	&	0.19&	0.40 	 $\pm$  	0.12 	 &	1.28 	 $\pm$  	0.12 	 &	22.86 	 $\pm$  	1.24 	 &	1.16 	 $\pm$  	0.07 	 &	22.63 	 $\pm$  	1.27&	1.38 	 $\pm$  	0.07 	 &	23.17 	 $\pm$  	1.18\tabularnewline  
56391.99 	&	00032785007	&	1.06&	0.52 	 $\pm$  	0.11 	 &	1.24 	 $\pm$  	0.11 	 &	13.81 	 $\pm$  	0.66 	 &	1.00 	 $\pm$  	0.06 	 &	13.74 	 $\pm$  	0.70&	1.23 	 $\pm$  	0.06 	 &	13.80 	 $\pm$  	0.65\tabularnewline  
56392.87 	&	00032785008	&	0.45&	0.49 	 $\pm$  	0.15 	 &	1.41 	 $\pm$  	0.14 	 &	7.80 	 $\pm$  	0.47 	 &	1.21 	 $\pm$  	0.08 	 &	7.61 	 $\pm$  	0.47&	1.43 	 $\pm$  	0.08 	 &	7.83 	 $\pm$  	0.44\tabularnewline  
56393.06 	&	00032785009	&	1.45&	0.51 	 $\pm$  	0.05 	 &	1.32 	 $\pm$  	0.05 	 &	19.09 	 $\pm$  	0.43 	 &	1.11 	 $\pm$  	0.03 	 &	18.75 	 $\pm$  	0.45&	1.32 	 $\pm$  	0.03 	 &	19.10 	 $\pm$  	0.42\tabularnewline  
56394.06 	&	00032785010	&	1.96&	0.45 	 $\pm$  	0.04 	 &	1.21 	 $\pm$  	0.04 	 &	29.70 	 $\pm$  	0.52 	 &	1.06 	 $\pm$  	0.02 	 &	29.42 	 $\pm$  	0.53&	1.26 	 $\pm$  	0.02 	 &	29.87 	 $\pm$  	0.50\tabularnewline  
56395.92 	&	00032785011	&	2.01&	0.34 	 $\pm$  	0.05 	 &	1.31 	 $\pm$  	0.05 	 &	12.25 	 $\pm$  	0.27 	 &	1.24 	 $\pm$  	0.03 	 &	12.16 	 $\pm$  	0.27&	1.46 	 $\pm$  	0.03 	 &	12.55 	 $\pm$  	0.25\tabularnewline  
56396.27 	&	00032785012	&	1.96&	0.44 	 $\pm$  	0.08 	 &	1.36 	 $\pm$  	0.08 	 &	5.49 	 $\pm$  	0.19 	 &	1.20 	 $\pm$  	0.04 	 &	5.39 	 $\pm$  	0.19&	1.41 	 $\pm$  	0.05 	 &	5.54 	 $\pm$  	0.17\tabularnewline  
56397.73 	&	00032785013	&	1.65&	-		 	 &	-			 &	-		     	 &	2.10 	 $\pm$  	0.16 	 &	0.46 	 $\pm$  	0.04&	2.36 	 $\pm$  	0.18 	 &	0.54 	 $\pm$  	0.05\tabularnewline  
56398.86 	&	00032785014	&	2.04&	-		 	 &	-			 &	-		     	 &	2.30 	 $\pm$  	0.20 	 &	0.34 	 $\pm$  	0.03&	2.58 	 $\pm$  	0.25 	 &	0.41 	 $\pm$  	0.04\tabularnewline  
56399.20 	&	00032785015	&	1.96&	-		 	 &	-			 &	-		     	 &	2.73 	 $\pm$  	0.25 	 &	0.32 	 $\pm$  	0.03&	3.21 	 $\pm$  	0.30 	 &	0.44 	 $\pm$  	0.06\tabularnewline 	    
\hline

\end{tabular}
\end{sidewaystable}

\begin{sidewaystable}
\centering
\caption{SAX J1748.9$-$2021}

\begin{tabular}{p{2.1cm}p{1.0cm}p{1.5cm}P{5.9cm}P{3.6cm}P{3.6cm}}
\tabularnewline
\hline
&&&$N_\text{H}$ free&  $N_\text{H}$ =  0.57 $\times10^{22}$ cm$^{-2}$&  $N_\text{H}$ =  1.41 $\times10^{22}$ cm$^{-2}$\tabularnewline
\cmidrule(r){4-4}\cmidrule(lr){5-5}\cmidrule(l){6-6}
\end{tabular}

\begin{tabular}{cccccccccc}
MJD& Obs ID & Exposure& $N_\text{H}$ & $\Gamma$ & L$_X$  &  $\Gamma$ & L$_X$&  $\Gamma$ & L$_X$\tabularnewline
(days)& & Time (ks)&(10$^{22}$ cm$^{-2}$)&&(10$^{35}$ erg s$^{-1}$)&&(10$^{35}$ erg s$^{-1}$)& &(10$^{35}$ erg s$^{-1}$)\tabularnewline
\hline

57071.78 	&	00033646001	&	1.97&	1.16 	 $\pm$  	0.13 	 &	1.26 	 $\pm$  	0.09 	 &	58.26 	 $\pm$  	1.96 	 &	0.89 	 $\pm$  	0.04 	 &	56.10 	 $\pm$  	1.91&	1.39 	 $\pm$  	0.05 	 &	60.32 	 $\pm$  	1.67\tabularnewline  
57083.97 	&	00033646005	&	1.06&	1.43 	 $\pm$  	0.03 	 &	1.77 	 $\pm$  	0.02 	 &	511.88 	 $\pm$  	6.24 	 &	1.19 	 $\pm$  	0.01 	 &	430.65 	 $\pm$  	3.49&	1.76 	 $\pm$  	0.01 	 &	508.09 	 $\pm$  	3.29\tabularnewline  
57085.70 	&	00033646007	&	0.97&	1.60 	 $\pm$  	0.04 	 &	1.86 	 $\pm$  	0.03 	 &	560.73 	 $\pm$  	8.14 	 &	1.19 	 $\pm$  	0.01 	 &	448.94 	 $\pm$  	3.78&	1.75 	 $\pm$  	0.01 	 &	529.04 	 $\pm$  	3.56\tabularnewline  
57087.43 	&	00033646008	&	0.96&	1.55 	 $\pm$  	0.03 	 &	1.62 	 $\pm$  	0.02 	 &	698.81 	 $\pm$  	7.20 	 &	1.02 	 $\pm$  	0.01 	 &	605.76 	 $\pm$  	4.66&	1.55 	 $\pm$  	0.01 	 &	676.49 	 $\pm$  	4.14\tabularnewline  
57091.55 	&	00033646010	&	0.41&	1.48 	 $\pm$  	0.05 	 &	1.60 	 $\pm$  	0.03 	 &	797.76 	 $\pm$  11.34   	 &	1.04 	 $\pm$  	0.02 	 &	700.75 	 $\pm$  	7.67&	1.56 	 $\pm$  	0.02 	 &	785.11 	 $\pm$  	6.81\tabularnewline  
57097.22 	&	00033646011	&	0.94&	1.59 	 $\pm$  	0.04 	 &	1.73 	 $\pm$  	0.02 	 &	706.46 	 $\pm$  	8.33 	 &	1.09 	 $\pm$  	0.01 	 &	591.33 	 $\pm$  	4.59&	1.63 	 $\pm$  	0.01 	 &	673.54 	 $\pm$  	4.20\tabularnewline  
57099.47 	&	00033646012	&	0.74&	1.54 	 $\pm$  	0.04 	 &	1.66 	 $\pm$  	0.02 	 &	706.59 	 $\pm$  	8.51 	 &	1.06 	 $\pm$  	0.01 	 &	606.75 	 $\pm$  	5.23&	1.59 	 $\pm$  	0.01 	 &	684.50 	 $\pm$  	4.70\tabularnewline  
57101.21 	&	00033646013	&	0.97&	1.42 	 $\pm$  	0.03 	 &	1.61 	 $\pm$  	0.02 	 &	661.06 	 $\pm$  	6.75 	 &	1.08 	 $\pm$  	0.01 	 &	581.18 	 $\pm$  	4.45&	1.60 	 $\pm$  	0.01 	 &	659.13 	 $\pm$  	4.03\tabularnewline  
57105.52 	&	00033646014	&	0.61&	1.33 	 $\pm$  	0.04 	 &	1.63 	 $\pm$  	0.03 	 &	605.06 	 $\pm$  	8.30 	 &	1.13 	 $\pm$  	0.01 	 &	534.53 	 $\pm$  	5.59&	1.67 	 $\pm$  	0.02 	 &	617.23 	 $\pm$  	5.12\tabularnewline  
57107.05 	&	00033646015	&	0.85&	1.55 	 $\pm$  	0.04 	 &	1.79 	 $\pm$  	0.03 	 &	553.54 	 $\pm$  	8.13 	 &	1.16 	 $\pm$  	0.01 	 &	456.15 	 $\pm$  	4.15&	1.71 	 $\pm$  	0.01 	 &	531.17 	 $\pm$  	3.87\tabularnewline  
57114.04 	&	00033646017	&	0.90&	1.37 	 $\pm$  	0.04 	 &	1.81 	 $\pm$  	0.03 	 &	438.86 	 $\pm$  	6.40 	 &	1.26 	 $\pm$  	0.01 	 &	367.99 	 $\pm$  	3.42&	1.83 	 $\pm$  	0.02 	 &	444.50 	 $\pm$  	3.33\tabularnewline  
57118.43 	&	00033646019	&	0.72&	1.41 	 $\pm$  	0.05 	 &	1.90 	 $\pm$  	0.03 	 &	370.02 	 $\pm$  	7.33 	 &	1.31 	 $\pm$  	0.02 	 &	299.72 	 $\pm$  	3.37&	1.90 	 $\pm$  	0.02 	 &	370.25 	 $\pm$  	3.43\tabularnewline  
57120.29 	&	00033646020	&	0.82&	1.48 	 $\pm$  	0.05 	 &	2.11 	 $\pm$  	0.04 	 &	316.92 	 $\pm$  	7.86 	 &	1.44 	 $\pm$  	0.02 	 &	234.74 	 $\pm$  	2.64&	2.06 	 $\pm$  	0.02 	 &	306.49 	 $\pm$  	2.99\tabularnewline  
57124.28 	&	00033646021	&	0.85&	1.32 	 $\pm$  	0.05 	 &	2.25 	 $\pm$  	0.05 	 &	176.02 	 $\pm$  	5.94 	 &	1.64 	 $\pm$  	0.02 	 &	127.94 	 $\pm$  	1.79&	2.32 	 $\pm$  	0.03 	 &	185.61 	 $\pm$  	2.57\tabularnewline  
57126.76 	&	00033646022	&	0.98&	0.85 	 $\pm$  	0.07 	 &	1.85 	 $\pm$  	0.07 	 &	51.99 	 $\pm$  	1.67 	 &	1.61 	 $\pm$  	0.03 	 &	47.64 	 $\pm$  	1.04&	2.25 	 $\pm$  	0.04 	 &	66.77 	 $\pm$  	1.42\tabularnewline  
57128.74 	&	00033646023	&	0.41&	1.53 	 $\pm$  	0.39 	 &	1.84 	 $\pm$  	0.26 	 &	24.68 	 $\pm$  	4.19 	 &	1.18 	 $\pm$  	0.12 	 &	20.27 	 $\pm$  	1.83&	1.76 	 $\pm$  	0.14 	 &	23.75 	 $\pm$  	1.68\tabularnewline  
57130.09 	&	00033646024	&	0.89&	1.45 	 $\pm$  	0.32 	 &	1.65 	 $\pm$  	0.20 	 &	17.13 	 $\pm$  	1.80 	 &	1.07 	 $\pm$  	0.09 	 &	15.00 	 $\pm$  	1.00&	1.62 	 $\pm$  	0.10 	 &	16.95 	 $\pm$  	0.90\tabularnewline  
57134.28 	&	00033646025	&	1.07&	1.43 	 $\pm$  	0.25 	 &	1.63 	 $\pm$  	0.17 	 &	14.42 	 $\pm$  	1.18 	 &	1.05 	 $\pm$  	0.08 	 &	12.78 	 $\pm$  	0.79&	1.62 	 $\pm$  	0.09 	 &	14.36 	 $\pm$  	0.70\tabularnewline  
57137.26 	&	00033646026	&	1.10&	1.43 	 $\pm$  	0.24 	 &	1.67 	 $\pm$  	0.17 	 &	11.11 	 $\pm$  	0.94 	 &	1.09 	 $\pm$  	0.08 	 &	9.74 	 $\pm$  	0.61&	1.66 	 $\pm$  	0.09 	 &	11.06 	 $\pm$  	0.54\tabularnewline  
57140.92 	&	00033646027	&	1.04&	1.41 	 $\pm$  	0.25 	 &	1.78 	 $\pm$  	0.18 	 &	10.33 	 $\pm$  	1.02 	 &	1.17 	 $\pm$  	0.09 	 &	8.83 	 $\pm$  	0.58&	1.78 	 $\pm$  	0.10 	 &	10.32 	 $\pm$  	0.52\tabularnewline  
57147.64 	&	00033646029	&	0.83&	1.60 	 $\pm$  	0.37 	 &	2.04 	 $\pm$  	0.26 	 &	8.31 	 $\pm$  	1.64 	 &	1.32 	 $\pm$  	0.12 	 &	6.32 	 $\pm$  	0.54&	1.92 	 $\pm$  	0.13 	 &	7.72 	 $\pm$  	0.52\tabularnewline  
57152.85 	&	00033646030	&	0.79&	1.54 	 $\pm$  	0.34 	 &	2.02 	 $\pm$  	0.24 	 &	9.99 	 $\pm$  	1.76 	 &	1.37 	 $\pm$  	0.11 	 &	7.65 	 $\pm$  	0.60&	1.94 	 $\pm$  	0.13 	 &	9.51 	 $\pm$  	0.61\tabularnewline  
57159.42 	&	00033646031	&	1.10&	1.62 	 $\pm$  	0.31 	 &	1.87 	 $\pm$  	0.21 	 &	8.51 	 $\pm$  	1.14 	 &	1.17 	 $\pm$  	0.10 	 &	6.88 	 $\pm$  	0.50&	1.74 	 $\pm$  	0.11 	 &	7.98 	 $\pm$  	0.45\tabularnewline  
57165.88 	&	00033646032	&	0.66&	1.24 	 $\pm$  	0.31 	 &	1.56 	 $\pm$  	0.22 	 &	11.84 	 $\pm$  	1.20 	 &	1.11 	 $\pm$  	0.11 	 &	10.83 	 $\pm$  	0.88&	1.67 	 $\pm$  	0.12 	 &	12.33 	 $\pm$  	0.78\tabularnewline  
57174.59 	&	00033646033	&	0.56&	1.30 	 $\pm$  	0.35 	 &	1.50 	 $\pm$  	0.24 	 &	18.25 	 $\pm$  	1.91 	 &	1.01 	 $\pm$  	0.11 	 &	16.91 	 $\pm$  	1.50&	1.57 	 $\pm$  	0.13 	 &	18.65 	 $\pm$  	1.28\tabularnewline  
57180.37 	&	00033646034	&	0.85&	1.36 	 $\pm$  	0.28 	 &	1.64 	 $\pm$  	0.19 	 &	18.61 	 $\pm$  	1.80 	 &	1.14 	 $\pm$  	0.09 	 &	16.40 	 $\pm$  	1.13&	1.67 	 $\pm$  	0.10 	 &	18.83 	 $\pm$  	1.03\tabularnewline  
57187.23 	&	00033646035	&	0.80&	1.44 	 $\pm$  	0.25 	 &	1.53 	 $\pm$  	0.17 	 &	27.98 	 $\pm$  	2.08 	 &	0.99 	 $\pm$  	0.08 	 &	25.36 	 $\pm$  	1.56&	1.51 	 $\pm$  	0.09 	 &	27.82 	 $\pm$  	1.36\tabularnewline  
57194.88 	&	00033646036	&	0.90&	1.50 	 $\pm$  	0.26 	 &	1.51 	 $\pm$  	0.16 	 &	33.05 	 $\pm$  	2.36 	 &	0.94 	 $\pm$  	0.08 	 &	29.92 	 $\pm$  	1.72&	1.46 	 $\pm$  	0.08 	 &	32.52 	 $\pm$  	1.50\tabularnewline  
57201.39 	&	00033646037	&	1.00&	1.59 	 $\pm$  	0.27 	 &	1.64 	 $\pm$  	0.17 	 &	19.41 	 $\pm$  	1.71 	 &	0.99 	 $\pm$  	0.08 	 &	16.91 	 $\pm$  	1.06&	1.54 	 $\pm$  	0.09 	 &	18.64 	 $\pm$  	0.93\tabularnewline  
57211.37 	&	00033646039	&	1.88&	1.50 	 $\pm$  	0.23 	 &	1.90 	 $\pm$  	0.16 	 &	8.18 	 $\pm$  	0.83 	 &	1.24 	 $\pm$  	0.08 	 &	6.61 	 $\pm$  	0.37&	1.84 	 $\pm$  	0.09 	 &	7.92 	 $\pm$  	0.35\tabularnewline  
57215.49 	&	00033646040	&	1.00&	1.54 	 $\pm$  	0.56 	 &	1.99 	 $\pm$  	0.41 	 &	2.84 	 $\pm$  	0.91 	 &	1.28 	 $\pm$  	0.18 	 &	2.27 	 $\pm$  	0.31&	1.90 	 $\pm$  	0.21 	 &	2.71 	 $\pm$  	0.27\tabularnewline

\hline
\end{tabular}
\end{sidewaystable}

\begin{sidewaystable}
\centering
\caption{IGR J17361$-$4441}

\begin{tabular}{p{2.1cm}p{1.0cm}p{1.5cm}P{5.9cm}P{3.5cm}P{3.5cm}}
\tabularnewline
\hline
&&&$N_\text{H}$ free&  $N_\text{H}$ =  0.25 $\times10^{22}$ cm$^{-2}$&  $N_\text{H}$ =  0.26 $\times10^{22}$ cm$^{-2}$\tabularnewline
\cmidrule(r){4-4}\cmidrule(lr){5-5}\cmidrule(l){6-6}
\end{tabular}

\begin{tabular}{cccccccccc}
MJD& Obs ID & Exposure& $N_\text{H}$ & $\Gamma$ & L$_X$  &  $\Gamma$ & L$_X$&  $\Gamma$ & L$_X$\tabularnewline
(days)& & Time (ks)&(10$^{22}$ cm$^{-2}$)&&(10$^{35}$ erg s$^{-1}$)&&(10$^{35}$ erg s$^{-1}$)& &(10$^{35}$ erg s$^{-1}$)\tabularnewline
\hline

55789.71 	&	00032072001	&	1.94&	0.09 	 $\pm$  	0.12 	 &	0.56 	 $\pm$  	0.15 	 &	11.24 	 $\pm$  	1.03 	 &	0.70 	 $\pm$  	0.10 	 &	11.01 	 $\pm$  	0.96  &	0.71 	 $\pm$  	0.10 	 &	11.00 	 $\pm$  	0.96\tabularnewline  
55790.66 	&	00032072002	&	2.58&	0.04 	 $\pm$  	0.09 	 &	0.50 	 $\pm$  	0.11 	 &	11.17 	 $\pm$  	0.80 	 &	0.69 	 $\pm$  	0.08 	 &	10.85 	 $\pm$  	0.73  &	0.70 	 $\pm$  	0.08 	 &	10.84 	 $\pm$  	0.73\tabularnewline  
55791.05 	&	00032072003	&	2.67&	0.22 	 $\pm$  	0.12 	 &	0.58 	 $\pm$  	0.12 	 &	11.17 	 $\pm$  	0.78 	 &	0.60 	 $\pm$  	0.08 	 &	11.13 	 $\pm$  	0.75  &	0.61 	 $\pm$  	0.08 	 &	11.12 	 $\pm$  	0.75\tabularnewline  
55792.39 	&	00032072004	&	3.22&	0.14 	 $\pm$  	0.09 	 &	0.56 	 $\pm$  	0.10 	 &	11.27 	 $\pm$  	0.70 	 &	0.64 	 $\pm$  	0.07 	 &	11.13 	 $\pm$  	0.66  &	0.65 	 $\pm$  	0.07 	 &	11.12 	 $\pm$  	0.66\tabularnewline  
55793.13 	&	00032072005	&	1.91&	0.36 	 $\pm$  	0.16 	 &	0.77 	 $\pm$  	0.15 	 &	11.28 	 $\pm$  	0.86 	 &	0.68 	 $\pm$  	0.09 	 &	11.38 	 $\pm$  	0.87  &	0.69 	 $\pm$  	0.09 	 &	11.36 	 $\pm$  	0.87\tabularnewline  
55796.75 	&	00032072006	&	2.48&	0.27 	 $\pm$  	0.13 	 &	0.74 	 $\pm$  	0.13 	 &	9.70 	 $\pm$  	0.70 	 &	0.73 	 $\pm$  	0.08 	 &	9.71 	 $\pm$  	0.70  &	0.73 	 $\pm$  	0.08 	 &	9.70 	 $\pm$  	0.69\tabularnewline  
55797.29 	&	00032072007	&	2.68&	0.23 	 $\pm$  	0.13 	 &	0.63 	 $\pm$  	0.13 	 &	10.14 	 $\pm$  	0.73 	 &	0.64 	 $\pm$  	0.08 	 &	10.13 	 $\pm$  	0.71  &	0.65 	 $\pm$  	0.08 	 &	10.12 	 $\pm$  	0.71\tabularnewline  
55798.15 	&	00032072008	&	3.02&	0.19 	 $\pm$  	0.11 	 &	0.62 	 $\pm$  	0.11 	 &	11.24 	 $\pm$  	0.73 	 &	0.66 	 $\pm$  	0.07 	 &	11.17 	 $\pm$  	0.70  &	0.67 	 $\pm$  	0.07 	 &	11.16 	 $\pm$  	0.70\tabularnewline  
55799.08 	&	00032072009	&	2.98&	0.26 	 $\pm$  	0.11 	 &	0.77 	 $\pm$  	0.12 	 &	10.38 	 $\pm$  	0.66 	 &	0.75 	 $\pm$  	0.07 	 &	10.40 	 $\pm$  	0.66  &	0.76 	 $\pm$  	0.07 	 &	10.39 	 $\pm$  	0.65\tabularnewline  
55800.08 	&	00032072010	&	2.98&	0.29 	 $\pm$  	0.12 	 &	0.74 	 $\pm$  	0.12 	 &	11.38 	 $\pm$  	0.73 	 &	0.71 	 $\pm$  	0.07 	 &	11.42 	 $\pm$  	0.73  &	0.71 	 $\pm$  	0.07 	 &	11.40 	 $\pm$  	0.73\tabularnewline  
55801.63 	&	00032072011	&	3.05&	0.29 	 $\pm$  	0.11 	 &	0.85 	 $\pm$  	0.12 	 &	9.87 	 $\pm$  	0.60 	 &	0.82 	 $\pm$  	0.07 	 &	9.89 	 $\pm$  	0.61  &	0.83 	 $\pm$  	0.07 	 &	9.89 	 $\pm$  	0.60\tabularnewline  
55802.09 	&	00032072012	&	3.90&	0.33 	 $\pm$  	0.11 	 &	0.87 	 $\pm$  	0.10 	 &	10.59 	 $\pm$  	0.55 	 &	0.80 	 $\pm$  	0.06 	 &	10.64 	 $\pm$  	0.56  &	0.81 	 $\pm$  	0.06 	 &	10.63 	 $\pm$  	0.56\tabularnewline  
55804.36 	&	00032072013	&	0.86&	0.21 	 $\pm$  	0.23 	 &	0.87 	 $\pm$  	0.26 	 &	9.58 	 $\pm$  	1.37 	 &	0.91 	 $\pm$  	0.15 	 &	9.55 	 $\pm$  	1.32  &	0.92 	 $\pm$  	0.15 	 &	9.54 	 $\pm$  	1.32\tabularnewline  
55806.24 	&	00032072014	&	1.02&	0.37 	 $\pm$  	0.22 	 &	0.93 	 $\pm$  	0.21 	 &	10.06 	 $\pm$  	1.09 	 &	0.83 	 $\pm$  	0.12 	 &	10.15 	 $\pm$  	1.12  &	0.84 	 $\pm$  	0.12 	 &	10.14 	 $\pm$  	1.11\tabularnewline  
55808.37 	&	00032072015	&	1.76&	0.41 	 $\pm$  	0.17 	 &	0.98 	 $\pm$  	0.15 	 &	10.42 	 $\pm$  	0.79 	 &	0.84 	 $\pm$  	0.09 	 &	10.49 	 $\pm$  	0.82  &	0.85 	 $\pm$  	0.09 	 &	10.48 	 $\pm$  	0.82\tabularnewline  
55810.38 	&	00032072016	&	3.09&	0.51 	 $\pm$  	0.19 	 &	1.08 	 $\pm$  	0.17 	 &	9.03 	 $\pm$  	0.65 	 &	0.87 	 $\pm$  	0.09 	 &	9.07 	 $\pm$  	0.70  &	0.88 	 $\pm$  	0.09 	 &	9.06 	 $\pm$  	0.70\tabularnewline  
55814.27 	&	00032072017	&	0.41&	0.62 	 $\pm$  	0.18 	 &	1.12 	 $\pm$  	0.15 	 &	10.89 	 $\pm$  	0.66 	 &	0.78 	 $\pm$  	0.19 	 &	15.25 	 $\pm$  	2.34  &	0.79 	 $\pm$  	0.18 	 &	15.24 	 $\pm$  	2.33\tabularnewline  
55814.27 	&	00032072017	&	2.41&	-		 	 &	-			 &	-		     	 &	0.82 	 $\pm$  	0.07 	 &	10.91 	 $\pm$  	0.72  &	0.83 	 $\pm$  	0.07 	 &	10.90 	 $\pm$  	0.72\tabularnewline  
55816.22 	&	00032072018	&	2.43&	0.47 	 $\pm$  	0.15 	 &	0.98 	 $\pm$  	0.14 	 &	10.47 	 $\pm$  	0.67 	 &	0.79 	 $\pm$  	0.08 	 &	10.56 	 $\pm$  	0.71  &	0.80 	 $\pm$  	0.08 	 &	10.55 	 $\pm$  	0.71\tabularnewline  
55818.28 	&	00032072019	&	2.24&	1.09 	 $\pm$  	0.29 	 &	1.59 	 $\pm$  	0.19 	 &	11.32 	 $\pm$  	1.01 	 &	1.03 	 $\pm$  	0.09 	 &	10.15 	 $\pm$  	0.70  &	1.04 	 $\pm$  	0.09 	 &	10.15 	 $\pm$  	0.70\tabularnewline  
55821.14 	&	00032072020	&	2.44&	0.57 	 $\pm$  	0.16 	 &	1.33 	 $\pm$  	0.15 	 &	8.87 	 $\pm$  	0.56 	 &	1.03 	 $\pm$  	0.08 	 &	8.75 	 $\pm$  	0.59  &	1.04 	 $\pm$  	0.08 	 &	8.75 	 $\pm$  	0.59\tabularnewline  
55824.22 	&	00032072021	&	2.03&	0.60 	 $\pm$  	0.16 	 &	1.42 	 $\pm$  	0.17 	 &	8.22 	 $\pm$  	0.59 	 &	1.08 	 $\pm$  	0.09 	 &	8.09 	 $\pm$  	0.63  &	1.09 	 $\pm$  	0.09 	 &	8.08 	 $\pm$  	0.63\tabularnewline  
55827.03 	&	00032072022	&	2.92&	0.53 	 $\pm$  	0.12 	 &	1.43 	 $\pm$  	0.14 	 &	8.17 	 $\pm$  	0.48 	 &	1.15 	 $\pm$  	0.07 	 &	8.02 	 $\pm$  	0.50  &	1.16 	 $\pm$  	0.07 	 &	8.02 	 $\pm$  	0.50\tabularnewline  
55830.11 	&	00032072023	&	2.52&	-		 	 &	-			 &	-		     	 &	1.13 	 $\pm$  	0.08 	 &	7.90 	 $\pm$  	0.56  &	1.14 	 $\pm$  	0.08 	 &	7.90 	 $\pm$  	0.56\tabularnewline  
55834.19 	&	00032072024	&	4.14&	0.80 	 $\pm$  	0.14 	 &	1.81 	 $\pm$  	0.14 	 &	7.23 	 $\pm$  	0.50 	 &	1.25 	 $\pm$  	0.06 	 &	6.48 	 $\pm$  	0.36  &	1.26 	 $\pm$  	0.06 	 &	6.48 	 $\pm$  	0.36\tabularnewline  
55838.08 	&	00032072025	&	4.14&	0.84 	 $\pm$  	0.16 	 &	1.94 	 $\pm$  	0.16 	 &	6.33 	 $\pm$  	0.56 	 &	1.34 	 $\pm$  	0.07 	 &	5.43 	 $\pm$  	0.33  &	1.35 	 $\pm$  	0.07 	 &	5.43 	 $\pm$  	0.32\tabularnewline  
55842.16 	&	00032072026	&	3.68&	0.78 	 $\pm$  	0.15 	 &	1.91 	 $\pm$  	0.16 	 &	5.78 	 $\pm$  	0.49 	 &	1.35 	 $\pm$  	0.08 	 &	5.07 	 $\pm$  	0.34  &	1.37 	 $\pm$  	0.08 	 &	5.07 	 $\pm$  	0.34\tabularnewline  
55846.24 	&	00032072027	&	3.82&	0.73 	 $\pm$  	0.14 	 &	2.24 	 $\pm$  	0.18 	 &	5.68 	 $\pm$  	0.66 	 &	1.61 	 $\pm$  	0.09 	 &	4.59 	 $\pm$  	0.31  &	1.62 	 $\pm$  	0.09 	 &	4.60 	 $\pm$  	0.31\tabularnewline  
55850.25 	&	00032072028	&	2.76&	0.72 	 $\pm$  	0.15 	 &	2.14 	 $\pm$  	0.19 	 &	5.27 	 $\pm$  	0.61 	 &	1.56 	 $\pm$  	0.09 	 &	4.37 	 $\pm$  	0.33  &	1.57 	 $\pm$  	0.09 	 &	4.38 	 $\pm$  	0.32\tabularnewline  
55854.21 	&	00032072029	&	0.87&	0.66 	 $\pm$  	0.32 	 &	2.37 	 $\pm$  	0.44 	 &	4.37 	 $\pm$  	1.53 	 &	1.86 	 $\pm$  	0.20 	 &	3.47 	 $\pm$  	0.48  &	1.87 	 $\pm$  	0.20 	 &	3.48 	 $\pm$  	0.48\tabularnewline  
55858.47 	&	00032072030	&	0.90&	-		 	 &	-			 &	-		     	 &	1.65 	 $\pm$  	0.23 	 &	3.28 	 $\pm$  	0.62  &	1.67 	 $\pm$  	0.23 	 &	3.28 	 $\pm$  	0.62\tabularnewline  
55862.10 	&	00032072031	&	3.87&	0.70 	 $\pm$  	0.15 	 &	2.30 	 $\pm$  	0.21 	 &	3.85 	 $\pm$  	0.54 	 &	1.72 	 $\pm$  	0.10 	 &	3.04 	 $\pm$  	0.23  &	1.73 	 $\pm$  	0.10 	 &	3.05 	 $\pm$  	0.23\tabularnewline  
55866.16 	&	00032072032	&	3.13&	0.59 	 $\pm$  	0.14 	 &	2.21 	 $\pm$  	0.22 	 &	3.82 	 $\pm$  	0.49 	 &	1.74 	 $\pm$  	0.11 	 &	3.24 	 $\pm$  	0.26  &	1.75 	 $\pm$  	0.11 	 &	3.25 	 $\pm$  	0.26\tabularnewline  
55869.37 	&	00032072033	&	0.94&	0.71 	 $\pm$  	0.39 	 &	2.06 	 $\pm$  	0.45 	 &	3.37 	 $\pm$  	1.03 	 &	1.57 	 $\pm$  	0.22 	 &	2.86 	 $\pm$  	0.48  &	1.58 	 $\pm$  	0.22 	 &	2.87 	 $\pm$  	0.48\tabularnewline

\hline
\end{tabular}
\end{sidewaystable}

\begin{sidewaystable}
\centering
\caption{SAX J1808.4$-$3658\textsuperscript{*}}

\begin{tabular}{p{2.0cm}p{0.9cm}p{1.4cm}P{5.9cm}P{6.2cm}}
\tabularnewline
\hline
&&&$N_\text{H}$ free&  $N_\text{H}$ =  0.14 $\times10^{22}$ cm$^{-2}$\tabularnewline
\cmidrule(r){4-4}\cmidrule(lr){5-5}
\end{tabular}

\begin{tabular}{cccccccc}

MJD& Obs ID & Exposure& $N_\text{H}$ & $\Gamma$ & L$_X$  &  $\Gamma$ & L$_X$\tabularnewline
(days)& & Time (ks)&(10$^{22}$ cm$^{-2}$)&&(10$^{35}$ erg s$^{-1}$)&&(10$^{35}$ erg s$^{-1}$) \tabularnewline
\hline

53538.03 	&	00030034001	&	0.12&	0.23 	 $\pm$  	0.11 	 &	2.14 	 $\pm$  	0.17 	 &	8.58 	 $\pm$  	0.77 	 &	2.01 	 $\pm$  	0.09 	 &	8.15 	 $\pm$  	0.45 \tabularnewline  
53538.03 	&	00030034001	&	0.77&	0.12 	 $\pm$  	0.06 	 &	1.89 	 $\pm$  	0.14 	 &	6.42 	 $\pm$  	0.42 	 &	1.94 	 $\pm$  	0.08 	 &	6.49 	 $\pm$  	0.39 \tabularnewline  
53541.45 	&	00030034002	&	1.05&	0.09 	 $\pm$  	0.10 	 &	1.72 	 $\pm$  	0.21 	 &	1.77 	 $\pm$  	0.18 	 &	1.81 	 $\pm$  	0.12 	 &	1.79 	 $\pm$  	0.17 \tabularnewline  
53544.87 	&	00030034003	&	0.76&	0.18 	 $\pm$  	0.06 	 &	2.14 	 $\pm$  	0.10 	 &	4.01 	 $\pm$  	0.22 	 &	2.09 	 $\pm$  	0.06 	 &	3.91 	 $\pm$  	0.13 \tabularnewline  
53544.88 	&	00030034003	&	0.16&	0.00 	 $\pm$  	0.10 	 &	1.51 	 $\pm$  	0.24 	 &	2.19 	 $\pm$  	0.36 	 &	1.76 	 $\pm$  	0.18 	 &	2.24 	 $\pm$  	0.33 \tabularnewline  
53558.74 	&	00030034005	&	0.85&	0.06 	 $\pm$  	0.05 	 &	2.21 	 $\pm$  	0.09 	 &	3.79 	 $\pm$  	0.17 	 &	2.34 	 $\pm$  	0.05 	 &	4.04 	 $\pm$  	0.10 \tabularnewline  
53564.03 	&	00030034006	&	0.71&	0.31 	 $\pm$  	0.07 	 &	2.25 	 $\pm$  	0.10 	 &	3.92 	 $\pm$  	0.23 	 &	2.02 	 $\pm$  	0.05 	 &	3.52 	 $\pm$  	0.12 \tabularnewline  
53564.03 	&	00030034006	&	0.39&	0.10 	 $\pm$  	0.08 	 &	1.62 	 $\pm$  	0.16 	 &	1.68 	 $\pm$  	0.14 	 &	1.69 	 $\pm$  	0.10 	 &	1.70 	 $\pm$  	0.13 \tabularnewline  
53595.17 	&	00030034012	&	0.43&	0.28 	 $\pm$  	0.13 	 &	2.23 	 $\pm$  	0.21 	 &	1.68 	 $\pm$  	0.20 	 &	2.04 	 $\pm$  	0.11 	 &	1.54 	 $\pm$  	0.10 \tabularnewline  
53595.17 	&	00030034012	&	2.34&	0.14 	 $\pm$  	0.06 	 &	1.96 	 $\pm$  	0.13 	 &	1.63 	 $\pm$  	0.10 	 &	1.96 	 $\pm$  	0.07 	 &	1.63 	 $\pm$  	0.09 \tabularnewline  
53610.42 	&	00030034014	&	0.07&	0.08 	 $\pm$  	0.06 	 &	1.87 	 $\pm$  	0.14 	 &	0.93 	 $\pm$  	0.06 	 &	1.76 	 $\pm$  	0.35 	 &	1.22 	 $\pm$  	0.28 \tabularnewline  
53610.42 	&	00030034014	&	1.90&	0.05 	 $\pm$  	0.02 	 &	1.85 	 $\pm$  	0.02 	 &	39.17 	 $\pm$  	0.42 	 &	1.99 	 $\pm$  	0.08 	 &	0.95 	 $\pm$  	0.06 \tabularnewline  
54733.90 	&	00325827000	&	1.61&	0.07 	 $\pm$  	0.03 	 &	1.87 	 $\pm$  	0.05 	 &	28.71 	 $\pm$  	0.65 	 &	1.97 	 $\pm$  	0.01 	 &	40.66 	 $\pm$  	0.34 \tabularnewline  
54733.97 	&	00030034026	&	0.85&	0.09 	 $\pm$  	0.02 	 &	1.96 	 $\pm$  	0.03 	 &	31.22 	 $\pm$  	0.38 	 &	1.96 	 $\pm$  	0.03 	 &	29.63 	 $\pm$  	0.52 \tabularnewline  
54734.57 	&	00030034027	&	1.00&	0.10 	 $\pm$  	0.02 	 &	1.86 	 $\pm$  	0.03 	 &	29.87 	 $\pm$  	0.34 	 &	2.03 	 $\pm$  	0.01 	 &	32.02 	 $\pm$  	0.29 \tabularnewline  
54735.50 	&	00030034028	&	1.14&	0.10 	 $\pm$  	0.02 	 &	2.00 	 $\pm$  	0.03 	 &	32.49 	 $\pm$  	0.38 	 &	1.91 	 $\pm$  	0.01 	 &	30.35 	 $\pm$  	0.28 \tabularnewline  
54736.17 	&	00030034029	&	1.12&	0.25 	 $\pm$  	0.02 	 &	2.31 	 $\pm$  	0.03 	 &	41.74 	 $\pm$  	0.60 	 &	2.05 	 $\pm$  	0.01 	 &	33.15 	 $\pm$  	0.28 \tabularnewline  
54737.51 	&	00030034030	&	1.02&	0.29 	 $\pm$  	0.02 	 &	2.39 	 $\pm$  	0.03 	 &	46.26 	 $\pm$  	0.76 	 &	2.15 	 $\pm$  	0.01 	 &	38.60 	 $\pm$  	0.29 \tabularnewline  
54738.45 	&	00030034031	&	0.85&	0.17 	 $\pm$  	0.03 	 &	2.18 	 $\pm$  	0.05 	 &	34.18 	 $\pm$  	0.78 	 &	2.18 	 $\pm$  	0.01 	 &	41.22 	 $\pm$  	0.33 \tabularnewline  
54739.05 	&	00030034032	&	1.23&	0.17 	 $\pm$  	0.02 	 &	2.20 	 $\pm$  	0.03 	 &	29.43 	 $\pm$  	0.42 	 &	2.14 	 $\pm$  	0.02 	 &	33.56 	 $\pm$  	0.46 \tabularnewline  
54740.60 	&	00030034033	&	1.13&	0.17 	 $\pm$  	0.02 	 &	2.18 	 $\pm$  	0.03 	 &	28.47 	 $\pm$  	0.40 	 &	2.16 	 $\pm$  	0.01 	 &	28.88 	 $\pm$  	0.24 \tabularnewline  
54741.93 	&	00030034034	&	1.20&	0.42 	 $\pm$  	0.17 	 &	2.23 	 $\pm$  	0.32 	 &	1.15 	 $\pm$  	0.21 	 &	2.14 	 $\pm$  	0.01 	 &	27.97 	 $\pm$  	0.24 \tabularnewline  
54758.00 	&	00030034036	&	0.46&	0.36 	 $\pm$  	0.14 	 &	2.39 	 $\pm$  	0.31 	 &	0.24 	 $\pm$  	0.05 	 &	1.77 	 $\pm$  	0.16 	 &	1.01 	 $\pm$  	0.13 \tabularnewline  
54760.27 	&	00030034037	&	1.63&	0.38 	 $\pm$  	0.13 	 &	2.19 	 $\pm$  	0.24 	 &	1.26 	 $\pm$  	0.16 	 &	1.99 	 $\pm$  	0.16 	 &	0.21 	 $\pm$  	0.02 \tabularnewline  
54762.08 	&	00030034038	&	0.71&	0.30 	 $\pm$  	0.05 	 &	2.18 	 $\pm$  	0.07 	 &	5.61 	 $\pm$  	0.21 	 &	1.79 	 $\pm$  	0.12 	 &	1.12 	 $\pm$  	0.11 \tabularnewline  
54764.82 	&	00030034039	&	1.04&	0.35 	 $\pm$  	0.08 	 &	2.20 	 $\pm$  	0.11 	 &	4.57 	 $\pm$  	0.29 	 &	2.23 	 $\pm$  	0.27 	 &	0.09 	 $\pm$  	0.02 \tabularnewline  
54770.11 	&	00030034041	&	1.04&	0.12 	 $\pm$  	0.09 	 &	1.69 	 $\pm$  	0.20 	 &	2.84 	 $\pm$  	0.28 	 &	1.96 	 $\pm$  	0.04 	 &	5.12 	 $\pm$  	0.12 \tabularnewline  
54774.53 	&	00030034042	&	0.56&	0.36 	 $\pm$  	0.08 	 &	2.59 	 $\pm$  	0.19 	 &	0.25 	 $\pm$  	0.03 	 &	1.93 	 $\pm$  	0.06 	 &	4.08 	 $\pm$  	0.15 \tabularnewline  
54774.53 	&	00030034042	&	0.40&	0.34 	 $\pm$  	0.01 	 &	2.29 	 $\pm$  	0.02 	 &	62.31 	 $\pm$  	0.79 	 &	1.73 	 $\pm$  	0.12 	 &	2.85 	 $\pm$  	0.27 \tabularnewline  
54776.00 	&	00030034043	&	3.39&	0.42 	 $\pm$  	0.01 	 &	2.31 	 $\pm$  	0.02 	 &	71.36 	 $\pm$  	0.92 	 &	2.14 	 $\pm$  	0.10 	 &	0.21 	 $\pm$  	0.01 \tabularnewline  
54778.07 	&	00030034044	&	1.67&	0.43 	 $\pm$  	0.02 	 &	2.34 	 $\pm$  	0.03 	 &	67.47 	 $\pm$  	1.18 	 &	2.01 	 $\pm$  	0.43 	 &	0.04 	 $\pm$  	0.01 \tabularnewline  
57123.75 	&	00030034073	&	0.49&	0.02 	 $\pm$  	0.11 	 &	1.25 	 $\pm$  	0.20 	 &	21.50 	 $\pm$  	2.58 	 &	1.44 	 $\pm$  	0.12 	 &	21.41 	 $\pm$  	2.36 \tabularnewline  
57123.82 	&	00637765000	&	0.65&	0.32 	 $\pm$  	0.03 	 &	1.76 	 $\pm$  	0.04 	 &	36.60 	 $\pm$  	0.57 	 &	1.56 	 $\pm$  	0.02 	 &	34.86 	 $\pm$  	0.49 \tabularnewline  
57125.02 	&	00030034074	&	5.15&	0.14 	 $\pm$  	0.04 	 &	1.30 	 $\pm$  	0.07 	 &	19.26 	 $\pm$  	0.70 	 &	1.30 	 $\pm$  	0.04 	 &	19.26 	 $\pm$  	0.70 \tabularnewline  
57126.09 	&	00030034076	&	1.62&	0.37 	 $\pm$  	0.02 	 &	1.75 	 $\pm$  	0.03 	 &	25.58 	 $\pm$  	0.31 	 &	1.51 	 $\pm$  	0.01 	 &	24.15 	 $\pm$  	0.26 \tabularnewline  
57127.74 	&	00030034077	&	0.82&	0.40 	 $\pm$  	0.03 	 &	1.83 	 $\pm$  	0.04 	 &	20.49 	 $\pm$  	0.39 	 &	1.54 	 $\pm$  	0.02 	 &	19.01 	 $\pm$  	0.31 \tabularnewline  
57128.02 	&	00081453001	&	0.07&	0.36 	 $\pm$  	0.44 	 &	1.56 	 $\pm$  	0.49 	 &	11.51 	 $\pm$  	2.57 	 &	1.34 	 $\pm$  	0.25 	 &	11.12 	 $\pm$  	2.46 \tabularnewline  
57128.02 	&	00081453001	&	1.72&	0.36 	 $\pm$  	0.02 	 &	1.79 	 $\pm$  	0.03 	 &	20.72 	 $\pm$  	0.25 	 &	1.55 	 $\pm$  	0.01 	 &	19.50 	 $\pm$  	0.21 \tabularnewline

\hline
\multicolumn{8}{p{18cm}}{\textsuperscript{*}\scriptsize{The photon index versus luminosity values for SAX J1808.4$-$3658 using the best-fit $N_\text{H}$ were not calculated and therefore are not presented here (see Appendix \ref{sect_best_fit_app}).}}

\end{tabular}
\end{sidewaystable}

\begin{sidewaystable}
\centering
\caption{SAX J1808.4$-$3658\textsuperscript{*} (continued)}

\begin{tabular}{p{2.0cm}p{0.9cm}p{1.4cm}P{5.9cm}P{6.2cm}}
\tabularnewline
\hline
&&&$N_\text{H}$ free&  $N_\text{H}$ =  0.14 $\times10^{22}$ cm$^{-2}$\tabularnewline
\cmidrule(r){4-4}\cmidrule(lr){5-5}
\end{tabular}

\begin{tabular}{cccccccc}

MJD& Obs ID & Exposure& $N_\text{H}$ & $\Gamma$ & L$_X$  &  $\Gamma$ & L$_X$\tabularnewline
(days)& & Time (ks)&(10$^{22}$ cm$^{-2}$)&&(10$^{35}$ erg s$^{-1}$)&&(10$^{35}$ erg s$^{-1}$) \tabularnewline
\hline

57129.41 	&	00033737001	&	5.16	&	0.40 	 $\pm$  	0.01 	 &	1.87 	 $\pm$  	0.02 	 &	19.58 	 $\pm$  	0.15 	 &	1.58 	 $\pm$  	0.01 	 &	18.07 	 $\pm$  	0.12 \tabularnewline   
57130.23 	&	00030034078	&	1.83	&	0.47 	 $\pm$  	0.02 	 &	1.91 	 $\pm$  	0.03 	 &	18.82 	 $\pm$  	0.26 	 &	1.55 	 $\pm$  	0.02 	 &	16.95 	 $\pm$  	0.19 \tabularnewline   
57131.83 	&	00030034079	&	0.39	&	0.44 	 $\pm$  	0.06 	 &	1.88 	 $\pm$  	0.07 	 &	15.65 	 $\pm$  	0.54 	 &	1.55 	 $\pm$  	0.04 	 &	14.28 	 $\pm$  	0.41 \tabularnewline   
57132.01 	&	00030034080	&	0.07	&	0.40 	 $\pm$  	0.04 	 &	1.89 	 $\pm$  	0.06 	 &	16.43 	 $\pm$  	0.44 	 &	1.42 	 $\pm$  	0.56 	 &	11.71 	 $\pm$  	7.21 \tabularnewline   
57132.01 	&	00030034080	&	0.53	&	-		 	 &	-			 &	-		     	 &	1.60 	 $\pm$  	0.03 	 	 &	15.12 	 $\pm$  	0.34 	 						\tabularnewline   
57133.87 	&	00030034081	&	1.97	&	0.44 	 $\pm$  	0.03 	 &	1.95 	 $\pm$  	0.03 	 &	13.96 	 $\pm$  	0.23 	 &	1.61 	 $\pm$  	0.02 	 &	12.57 	 $\pm$  	0.16 \tabularnewline  
57134.75 	&	00030034082	&	2.17	&	0.39 	 $\pm$  	0.02 	 &	1.90 	 $\pm$  	0.03 	 &	13.22 	 $\pm$  	0.19 	 &	1.62 	 $\pm$  	0.02 	 &	12.15 	 $\pm$  	0.14 \tabularnewline  
57136.74 	&	00033737002	&	3.71	&	0.46 	 $\pm$  	0.02 	 &	2.05 	 $\pm$  	0.03 	 &	10.23 	 $\pm$  	0.15 	 &	1.69 	 $\pm$  	0.01 	 &	8.97 	 $\pm$  	0.09 \tabularnewline  
57141.65 	&	00033737003	&	2.52	&	0.52 	 $\pm$  	0.04 	 &	2.37 	 $\pm$  	0.06 	 &	4.63 	 $\pm$  	0.17 	 &	1.90 	 $\pm$  	0.03 	 &	3.66 	 $\pm$  	0.07 \tabularnewline  
57143.00 	&	00033737004	&	0.07	&	1.00 	 $\pm$  	0.67 	 &	2.97 	 $\pm$  	0.94 	 &	1.66 	 $\pm$  	2.68 	 &	1.86 	 $\pm$  	0.38 	 &	0.82 	 $\pm$  	0.24 \tabularnewline  
57143.00 	&	00033737004	&	5.36	&	0.54 	 $\pm$  	0.05 	 &	2.68 	 $\pm$  	0.08 	 &	1.36 	 $\pm$  	0.09 	 &	2.13 	 $\pm$  	0.04 	 &	0.97 	 $\pm$  	0.02 \tabularnewline  
57144.73 	&	00030034083	&	0.41	&	0.44 	 $\pm$  	0.17 	 &	2.16 	 $\pm$  	0.32 	 &	1.32 	 $\pm$  	0.21 	 &	1.64 	 $\pm$  	0.16 	 &	1.21 	 $\pm$  	0.16 \tabularnewline  
57145.85 	&	00030034084	&	1.11	&	0.39 	 $\pm$  	0.10 	 &	2.22 	 $\pm$  	0.19 	 &	0.82 	 $\pm$  	0.08 	 &	1.81 	 $\pm$  	0.10 	 &	0.72 	 $\pm$  	0.05 \tabularnewline  
57146.31 	&	00030034085	&	0.99	&	0.46 	 $\pm$  	0.12 	 &	2.43 	 $\pm$  	0.23 	 &	0.63 	 $\pm$  	0.09 	 &	2.24 	 $\pm$  	0.12 	 &	0.57 	 $\pm$  	0.04 \tabularnewline  
57147.06 	&	00030034086	&	0.68	&	0.46 	 $\pm$  	0.21 	 &	2.60 	 $\pm$  	0.38 	 &	0.43 	 $\pm$  	0.13 	 &	2.43 	 $\pm$  	0.19 	 &	0.39 	 $\pm$  	0.04 \tabularnewline  
57148.78 	&	00030034087	&	1.09	&	0.41 	 $\pm$  	0.13 	 &	2.57 	 $\pm$  	0.28 	 &	0.44 	 $\pm$  	0.08 	 &	2.47 	 $\pm$  	0.15 	 &	0.41 	 $\pm$  	0.03 \tabularnewline  
57151.57 	&	00033737005	&	5.82	&	0.46 	 $\pm$  	0.05 	 &	2.53 	 $\pm$  	0.11 	 &	0.49 	 $\pm$  	0.03 	 &	2.33 	 $\pm$  	0.06 	 &	0.44 	 $\pm$  	0.01 \tabularnewline  
57151.97 	&	00030034088	&	0.96	&	0.53 	 $\pm$  	0.16 	 &	2.76 	 $\pm$  	0.35 	 &	0.50 	 $\pm$  	0.14 	 &	2.42 	 $\pm$  	0.17 	 &	0.41 	 $\pm$  	0.04 \tabularnewline  
57152.17 	&	00030034090	&	1.03	&	0.36 	 $\pm$  	0.09 	 &	2.23 	 $\pm$  	0.18 	 &	0.85 	 $\pm$  	0.08 	 &	2.21 	 $\pm$  	0.10 	 &	0.84 	 $\pm$  	0.05 \tabularnewline  
57152.64 	&	00030034089	&	1.06	&	0.13 	 $\pm$  	0.08 	 &	1.72 	 $\pm$  	0.16 	 &	3.62 	 $\pm$  	0.28 	 &	2.08 	 $\pm$  	0.10 	 &	3.96 	 $\pm$  	0.25 \tabularnewline  
57154.56 	&	00030034091	&	1.14	&	0.09 	 $\pm$  	0.06 	 &	1.69 	 $\pm$  	0.13 	 &	4.61 	 $\pm$  	0.30 	 &	2.13 	 $\pm$  	0.09 	 &	5.19 	 $\pm$  	0.27 \tabularnewline  
57155.69 	&	00030034092	&	0.95	&	0.10 	 $\pm$  	0.05 	 &	1.93 	 $\pm$  	0.13 	 &	6.24 	 $\pm$  	0.37 	 &	2.41 	 $\pm$  	0.08 	 &	7.49 	 $\pm$  	0.33 \tabularnewline  
57157.77 	&	00033737006	&	3.68	&	0.51 	 $\pm$  	0.05 	 &	2.37 	 $\pm$  	0.10 	 &	1.27 	 $\pm$  	0.07 	 &	2.11 	 $\pm$  	0.05 	 &	1.12 	 $\pm$  	0.04 \tabularnewline  
57161.82 	&	00030034094	&	0.97	&	0.14 	 $\pm$  	0.06 	 &	1.83 	 $\pm$  	0.14 	 &	4.31 	 $\pm$  	0.29 	 &	2.19 	 $\pm$  	0.09 	 &	4.83 	 $\pm$  	0.26 \tabularnewline  
57164.67 	&	00030034095	&	4.17	&	0.09 	 $\pm$  	0.02 	 &	2.15 	 $\pm$  	0.03 	 &	7.18 	 $\pm$  	0.10 	 &	2.53 	 $\pm$  	0.02 	 &	8.87 	 $\pm$  	0.07 \tabularnewline  
57168.60 	&	00030034096	&	0.14	&	0.28 	 $\pm$  	0.20 	 &	1.92 	 $\pm$  	0.36 	 &	2.18 	 $\pm$  	0.36 	 &	2.02 	 $\pm$  	0.21 	 &	2.25 	 $\pm$  	0.30 \tabularnewline  
57169.08 	&	00033801002	&	0.13	&	0.13 	 $\pm$  	0.20 	 &	1.77 	 $\pm$  	0.37 	 &	1.19 	 $\pm$  	0.20 	 &	2.09 	 $\pm$  	0.22 	 &	1.31 	 $\pm$  	0.18 \tabularnewline  
57171.45 	&	00030034097	&	5.55&	-		 	 &	-			 &	-		     	 	 &	2.57 	 $\pm$  	0.17 	 &	0.06 	 $\pm$  	0.01 							\tabularnewline  
57172.78 	&	00033801003	&	1.10	&	0.56 	 $\pm$  	0.11 	 &	2.68 	 $\pm$  	0.21 	 &	0.92 	 $\pm$  	0.15 	 &	2.33 	 $\pm$  	0.11 	 &	0.74 	 $\pm$  	0.05 \tabularnewline  
57175.44 	&	00033801004	&	1.16&	-		 	 &	-			 &	-		     	 	 &	2.42 	 $\pm$  	0.49 	 &	0.05 	 $\pm$  	0.01 							\tabularnewline  
57179.31 	&	00030034098	&	5.37	&	0.22 	 $\pm$  	0.04 	 &	2.06 	 $\pm$  	0.08 	 &	0.58 	 $\pm$  	0.02 	 &	2.29 	 $\pm$  	0.05 	 &	0.63 	 $\pm$  	0.02 \tabularnewline  
57182.16 	&	00033801008	&	2.10&	-		 	 &	-			 &	-		     		 &	3.09 	 $\pm$  	0.32 	 &	0.07 	 $\pm$  	0.01 							\tabularnewline  
57190.01 	&	00033801011	&	1.80&	-		 	 &	-			 &	-		     		 &	2.77 	 $\pm$  	0.19 	 &	0.16 	 $\pm$  	0.01 							\tabularnewline  
57192.01 	&	00030034101	&	5.21	&	0.43 	 $\pm$  	0.06 	 &	2.64 	 $\pm$  	0.14 	 &	0.40 	 $\pm$  	0.04 	 &	2.49 	 $\pm$  	0.07 	 &	0.37 	 $\pm$  	0.01 \tabularnewline  
57193.61 	&	00033801012	&	2.12&	-		 	 &	-			 &	-		     		 &	2.68 	 $\pm$  	0.23 	 &	0.10 	 $\pm$  	0.01 							\tabularnewline  
57208.58 	&	00033801016	&	1.75	&	0.53 	 $\pm$  	0.14 	 &	2.85 	 $\pm$  	0.31 	 &	0.44 	 $\pm$  	0.11 	 &	2.51 	 $\pm$  	0.15 	 &	0.35 	 $\pm$  	0.03 \tabularnewline  
57211.58 	&	00033801017	&	1.79&	-		 	 &	-			 &	-		     		 &	2.69 	 $\pm$  	0.40 	 &	0.05 	 $\pm$  	0.01 							\tabularnewline  
57241.37 	&	00030034108	&	5.84&	-		 	 &	-			 &	-		     		 &	2.51 	 $\pm$  	0.28 	 &	0.03 	 $\pm$  	0.00 							\tabularnewline  

\hline

\multicolumn{8}{p{18cm}}{\textsuperscript{*}\scriptsize{The photon index versus luminosity values for SAX J1808.4$-$3658 using the best-fit $N_\text{H}$ were not calculated and therefore are not presented here (see Appendix \ref{sect_best_fit_app}).}}
\end{tabular}
\end{sidewaystable}

%%%%%%%%%%%%%%%%%%%%%%%%%%%%%%%%%%%%%%%%%%%%%%%%%%
%%%%%%%%%%%%%%%%%%%%%%%%%%%%%%%%%%%%%%%%%%%%%%%%%%

% Don't change these lines
\bsp	% typesetting comment
\label{lastpage}

\begin{thebibliography}{}
\makeatletter
\relax
\def\mn@urlcharsother{\let\do\@makeother \do\$\do\&\do\#\do\^\do\_\do\%\do\~}
\def\mn@doi{\begingroup\mn@urlcharsother \@ifnextchar [ {\mn@doi@}
  {\mn@doi@[]}}
\def\mn@doi@[#1]#2{\def\@tempa{#1}\ifx\@tempa\@empty \href
  {http://dx.doi.org/#2} {doi:#2}\else \href {http://dx.doi.org/#2} {#1}\fi
  \endgroup}
\def\mn@eprint#1#2{\mn@eprint@#1:#2::\@nil}
\def\mn@eprint@arXiv#1{\href {http://arxiv.org/abs/#1} {{\tt arXiv:#1}}}
\def\mn@eprint@dblp#1{\href {http://dblp.uni-trier.de/rec/bibtex/#1.xml}
  {dblp:#1}}
\def\mn@eprint@#1:#2:#3:#4\@nil{\def\@tempa {#1}\def\@tempb {#2}\def\@tempc
  {#3}\ifx \@tempc \@empty \let \@tempc \@tempb \let \@tempb \@tempa \fi \ifx
  \@tempb \@empty \def\@tempb {arXiv}\fi \@ifundefined
  {mn@eprint@\@tempb}{\@tempb:\@tempc}{\expandafter \expandafter \csname
  mn@eprint@\@tempb\endcsname \expandafter{\@tempc}}}

\bibitem[\protect\citeauthoryear{Altamirano, Casella, Patruno, Wijnands  \& Van
  Der~Klis}{Altamirano et~al.}{2008}]{altamirano2008intermittent}
Altamirano D.,  Casella P.,  Patruno A.,  Wijnands R.,   Van Der~Klis M.,
  2008, ApJ Letters, 674, L45

\bibitem[\protect\citeauthoryear{Archibald et~al.,}{Archibald
  et~al.}{2009}]{archibald2009radio}
Archibald A.~M.,  et~al., 2009, Science, 324, 1411

\bibitem[\protect\citeauthoryear{Armas~Padilla, Degenaar, Patruno, Russell,
  Linares, Maccarone, Homan  \& Wijnands}{Armas~Padilla
  et~al.}{2011}]{padilla2011x}
Armas~Padilla M.,  Degenaar N.,  Patruno A.,  Russell D.,  Linares M.,
  Maccarone T.,  Homan J.,   Wijnands R.,  2011, MNRAS, 417, 659

\bibitem[\protect\citeauthoryear{Bahramian et~al.,}{Bahramian
  et~al.}{2013}]{bahramian2013discovery}
Bahramian A.,  et~al., 2013, ApJ, 780, 127

\bibitem[\protect\citeauthoryear{Bellini et~al.,}{Bellini
  et~al.}{2013}]{bellini2013intriguing}
Bellini A.,  et~al., 2013, ApJ, 765, 32

\bibitem[\protect\citeauthoryear{Bozzo et~al.,}{Bozzo
  et~al.}{2011}]{bozzo2011igr}
Bozzo E.,  et~al., 2011, A\&A, 535, L1

\bibitem[\protect\citeauthoryear{Bozzo, Papitto, Ferrigno  \& Belloni}{Bozzo
  et~al.}{2014}]{bozzo2014100}
Bozzo E.,  Papitto A.,  Ferrigno C.,   Belloni T.,  2014, A\&A, 570, L2

\bibitem[\protect\citeauthoryear{Bozzo, Kuulkers  \& Ferrigno}{Bozzo
  et~al.}{2015}]{bozzo2015swift}
Bozzo E.,  Kuulkers E.,   Ferrigno C.,  2015, ATel, 7106, 1

\bibitem[\protect\citeauthoryear{Chenevez et~al.,}{Chenevez
  et~al.}{2012}]{chenevez2012integral}
Chenevez J.,  et~al., 2012, ATel, 4050, 1

\bibitem[\protect\citeauthoryear{Dalessandro, Lanzoni, Ferraro, Rood, Milone,
  Piotto  \& Valenti}{Dalessandro et~al.}{2008}]{dalessandro2008blue}
Dalessandro E.,  Lanzoni B.,  Ferraro F.,  Rood R.,  Milone A.,  Piotto G.,
  Valenti E.,  2008, ApJ, 677, 1069

\bibitem[\protect\citeauthoryear{Degenaar \& Wijnands}{Degenaar \&
  Wijnands}{2012}]{degenaar2012strong}
Degenaar N.,  Wijnands R.,  2012, MNRAS, 422, 581

\bibitem[\protect\citeauthoryear{Degenaar, Wijnands, Cackett, Homan, Kuulkers,
  Maccarone, van~der Klis  et~al.}{Degenaar et~al.}{2012}]{degenaar2012four}
Degenaar N.,  Wijnands R.,  Cackett E.,  Homan J.,  Kuulkers E.,  Maccarone T.,
   van~der Klis M.,   et~al., 2012, A\&A, 545, A49

\bibitem[\protect\citeauthoryear{Degenaar et~al.,}{Degenaar
  et~al.}{2016}]{degenaar2016disk}
Degenaar N.,  et~al., 2016, MNRAS, 461, 4049

\bibitem[\protect\citeauthoryear{Del~Santo, Nucita, Lodato, Manni, De~Paolis,
  Farihi, De~Cesare  \& Segreto}{Del~Santo et~al.}{2014}]{del2014puzzling}
Del~Santo M.,  Nucita A.~A.,  Lodato G.,  Manni L.,  De~Paolis F.,  Farihi J.,
  De~Cesare G.,   Segreto A.,  2014, MNRAS, 444, 93

\bibitem[\protect\citeauthoryear{Dickey \& Lockman}{Dickey \&
  Lockman}{1990}]{dickey1990hi}
Dickey J.~M.,  Lockman F.~J.,  1990, ARA\&A, 28, 215

\bibitem[\protect\citeauthoryear{Ferrigno, Bozzo, Rodriguez  \&
  Gibaud}{Ferrigno et~al.}{2011}]{ferrigno2011swift}
Ferrigno C.,  Bozzo E.,  Rodriguez J.,   Gibaud L.,  2011, ATel, 3566, 1

\bibitem[\protect\citeauthoryear{Ferrigno et~al.,}{Ferrigno
  et~al.}{2014}]{ferrigno2014hiccup}
Ferrigno C.,  et~al., 2014, A\&A, 567, A77

\bibitem[\protect\citeauthoryear{Fridriksson, Homan  \& Remillard}{Fridriksson
  et~al.}{2015}]{fridriksson2015common}
Fridriksson J.~K.,  Homan J.,   Remillard R.~A.,  2015, ApJ, 809, 52

\bibitem[\protect\citeauthoryear{Galloway \& Cumming}{Galloway \&
  Cumming}{2006}]{galloway2006helium}
Galloway D.~K.,  Cumming A.,  2006, ApJ, 652, 559

\bibitem[\protect\citeauthoryear{G{\"u}ver \& {\"O}zel}{G{\"u}ver \&
  {\"O}zel}{2009}]{guver2009relation}
G{\"u}ver T.,  {\"O}zel F.,  2009, MNRAS, 400, 2050

\bibitem[\protect\citeauthoryear{Harris}{Harris}{1996}]{harris1996catalog}
Harris W.~E.,  1996, The Astronomical Journal, 112, 1487

\bibitem[\protect\citeauthoryear{Homan et~al.,}{Homan
  et~al.}{2010}]{homan2010xte}
Homan J.,  et~al., 2010, ApJ, 719, 201

\bibitem[\protect\citeauthoryear{Lewis et~al.,}{Lewis
  et~al.}{2010}]{lewis2010double}
Lewis F.,  et~al., 2010, A\&A, 517, A72

\bibitem[\protect\citeauthoryear{Linares et~al.,}{Linares
  et~al.}{2014}]{linares2014neutron}
Linares M.,  et~al., 2014, MNRAS, 438, 251

\bibitem[\protect\citeauthoryear{Ludlam et~al.,}{Ludlam
  et~al.}{2016}]{ludlam2016nustar}
Ludlam R.,  et~al., 2016, ApJ, 824, 37

\bibitem[\protect\citeauthoryear{Ortolani, Barbuy  \& Bica}{Ortolani
  et~al.}{1994}]{ortolani1994low}
Ortolani S.,  Barbuy B.,   Bica E.,  1994, A\&A Supplement Series, 108

\bibitem[\protect\citeauthoryear{Ortolani, Barbuy, Bica, Zoccali  \&
  Renzini}{Ortolani et~al.}{2007}]{ortolani2007distances}
Ortolani S.,  Barbuy B.,  Bica E.,  Zoccali M.,   Renzini A.,  2007, A\&A, 470,
  1043

\bibitem[\protect\citeauthoryear{Papitto et~al.,}{Papitto
  et~al.}{2013}]{papitto2013swings}
Papitto A.,  et~al., 2013, Nature, 501, 517

\bibitem[\protect\citeauthoryear{Parikh et~al.,}{Parikh
  et~al.}{2016}]{parikh2016potential}
Parikh A.,  et~al., 2016, arXiv preprint arXiv:1609.06703

\bibitem[\protect\citeauthoryear{Patruno, Altamirano, Hessels, Casella,
  Wijnands  \& van~der Klis}{Patruno et~al.}{2009}]{patruno2009phase}
Patruno A.,  Altamirano D.,  Hessels J.~W.,  Casella P.,  Wijnands R.,
  van~der Klis M.,  2009, ApJ, 690, 1856

\bibitem[\protect\citeauthoryear{Patruno et~al.,}{Patruno
  et~al.}{2016}]{patruno2016radio}
Patruno A.,  et~al., 2016, arXiv preprint arXiv:1611.06023

\bibitem[\protect\citeauthoryear{Pintore et~al.,}{Pintore
  et~al.}{2016}]{pintore2016broad}
Pintore F.,  et~al., 2016, MNRAS, 457, 2988

\bibitem[\protect\citeauthoryear{Plotkin, Gallo  \& Jonker}{Plotkin
  et~al.}{2013}]{plotkin2013x}
Plotkin R.~M.,  Gallo E.,   Jonker P.~G.,  2013, ApJ, 773, 59

\bibitem[\protect\citeauthoryear{Plotkin et~al.,}{Plotkin
  et~al.}{2016}]{plotkin20162015}
Plotkin R.,  et~al., 2016, arXiv preprint arXiv:1611.02810

\bibitem[\protect\citeauthoryear{Predehl \& Schmitt}{Predehl \&
  Schmitt}{1995}]{predehl1995x}
Predehl P.,  Schmitt J.~H.,  1995, A\& A, 293

\bibitem[\protect\citeauthoryear{Reynolds, Reis, Miller, Cackett  \&
  Degenaar}{Reynolds et~al.}{2014}]{reynolds2014quiescent}
Reynolds M.~T.,  Reis R.~C.,  Miller J.~M.,  Cackett E.~M.,   Degenaar N.,
  2014, MNRAS, 441, 3656

\bibitem[\protect\citeauthoryear{Tetarenko et~al.,}{Tetarenko
  et~al.}{2016}]{tetarenko2016disc}
Tetarenko A.,  et~al., 2016, MNRAS, 460, 345

\bibitem[\protect\citeauthoryear{Verner, Ferland, Korista  \& Yakovlev}{Verner
  et~al.}{1996}]{verner1996atomic}
Verner D.,  Ferland G.,  Korista K.,   Yakovlev D.,  1996, ApJ, 465

\bibitem[\protect\citeauthoryear{Wachter, Leach  \& Kellogg}{Wachter
  et~al.}{1979}]{wachter1979parameter}
Wachter K.,  Leach R.,   Kellogg E.,  1979, ApJ, 230, 274

\bibitem[\protect\citeauthoryear{Wijnands, Heinke, Pooley, Edmonds, Lewin,
  Grindlay, Jonker  \& Miller}{Wijnands et~al.}{2005}]{wijnands2005hard}
Wijnands R.,  Heinke C.~O.,  Pooley D.,  Edmonds P.~D.,  Lewin W.~H.,  Grindlay
  J.~E.,  Jonker P.~G.,   Miller J.~M.,  2005, ApJ, 618, 883

\bibitem[\protect\citeauthoryear{Wijnands, Degenaar, Padilla, Altamirano,
  Cavecchi, Linares, Bahramian  \& Heinke}{Wijnands
  et~al.}{2015}]{wijnands2015low}
Wijnands R.,  Degenaar N.,  Padilla M.~A.,  Altamirano D.,  Cavecchi Y.,
  Linares M.,  Bahramian A.,   Heinke C.,  2015, MNRAS, 454, 1371

\bibitem[\protect\citeauthoryear{Wijnands, Parikh, Altamirano  \&
  Degenaar}{Wijnands et~al.}{in preparation}]{wijnands2017}
Wijnands R.,  Parikh A.,  Altamirano D.,   Degenaar N.,  in preparation

\bibitem[\protect\citeauthoryear{Wilms, Allen  \& McCray}{Wilms
  et~al.}{2000}]{wilms2000absorption}
Wilms J.,  Allen A.,   McCray R.,  2000, ApJ, 542, 914

\makeatother
\end{thebibliography}
\end{document}